\documentclass [11pt]{article}
 \pdfoutput=1
\usepackage{graphicx}
\usepackage{amsmath,amssymb} 
\usepackage{latexsym}
\usepackage{mathrsfs}
\usepackage{bbold}
\usepackage{color}
\usepackage{slashed}
\definecolor{red}{rgb}{1,0,0}
\usepackage{cancel}
\usepackage{comment}
\usepackage{cite,mathtools}

\definecolor{red}{rgb}{1,0,0}

\makeatletter
\def\section{\@startsection {section}{1}{\z@}{-3.5ex plus -1ex minus
 -.2ex}{2.3ex plus .2ex}{\large\bf}}
\def\subsection{\@startsection{subsection}{2}{\z@}{-3.25ex plus -1ex
minus -.2ex}{1.5ex plus .2ex}{\normalsize\bf}}
\makeatother
\makeatletter

\@addtoreset{equation}{section}

\makeatother

\def\bea{\begin{eqnarray}} \def\eea{\end{eqnarray}}
\def\be{\begin{equation}} \def\ee{\end{equation}} \def\nn{\nonumber}

\newcommand{\promille}{%
  \relax\ifmmode\promillezeichen
        \else\leavevmode\(\mathsurround=0pt\promillezeichen\)\fi}
\newcommand{\promillezeichen}{%
  \kern-.05em%
  \raise.5ex\hbox{\the\scriptfont0 0}%
  \kern-.15em/\kern-.15em%
  \lower.25ex\hbox{\the\scriptfont0 00}}

\usepackage[colorlinks,linkcolor=black,citecolor=blue,urlcolor=blue,linktocpage]{hyperref}

\begin{document}

\thispagestyle{empty}

\begin{center}

\begin{center}

\vspace*{1.1cm}

{\Large\bf The Effective Bootstrap}
\end{center}

\vspace{0.8cm}

{ \bf  Alejandro Castedo Echeverri$^{a}$, Benedict von Harling$^{b}$ and Marco Serone$^{a,c}$}\\

\vspace{1.cm}
${}^a\!\!$
{\em SISSA and INFN, Via Bonomea 265, I-34136 Trieste, Italy} \\
\vspace{.1cm}
${}^b\!\!$
{\em DESY, Notkestrasse 85, 22607 Hamburg, Germany} \\
\vspace{.1cm}
${}^c\!\!$
{\em ICTP, Strada Costiera 11, I-34151 Trieste, Italy}

\end{center}

\vspace{1cm}

\centerline{\bf Abstract}
\vspace{2 mm}
\begin{quote}

We study the numerical bounds obtained using a conformal-bootstrap method -- advocated in ref.\cite{Hogervorst:2013sma} but never implemented so far --
where different points in the plane of conformal cross ratios $z$ and $\bar z$ are sampled. 
In contrast to the most used method based on derivatives evaluated at the symmetric point $z=\bar z =1/2$, 
we can consistently ``integrate out" higher-dimensional operators and get a reduced simpler, and faster to solve, set of bootstrap equations.  
We test this ``effective" bootstrap by studying the 3D Ising and $O(n)$ vector models and bounds on generic 4D CFTs,
for which extensive results are already available in the literature. We also determine the scaling dimensions of certain scalar operators in the $O(n)$
 vector models, with $n=2,3,4$,  which have not yet been computed using bootstrap techniques.

\end{quote}

\newpage

\tableofcontents


\section{Introduction}

There has recently been a great revival of interest in the conformal bootstrap program \cite{Ferrara:1973yt,Polyakov:1974gs} after ref.\cite{Rattazzi:2008pe} observed that 
its applicability extends to Conformal Field Theories (CFTs) in $d>2$ dimensions. 
Since ref.\cite{Rattazzi:2008pe}, considerable progress has been achieved in understanding CFTs in $d\geq 2$ dimensions, both numerically and analytically.
Probably the most striking progress has been made in the numerical study of the 3D Ising model,
where amazingly precise operator dimensions and OPE coefficients have been determined
\cite{ElShowk:2012ht,El-Showk:2014dwa,Kos:2016ysd}.

Essentially all numerical bootstrap studies so far have used the constraints imposed by crossing symmetry on 4-point correlators evaluated at a specific 
value of the conformal cross-ratios, $u=v=1/4$, or equivalently in $z$-coordinates at $z=\bar z=1/2$ \cite{Dolan:2000ut}.
This is the point of best convergence for the combined conformal block expansions in the $s$ and $t$ channels.
Taking higher and higher derivatives of the bootstrap equations evaluated at this point has proven to be very effective and successful in obtaining increasingly better bounds.
We will denote this method in the following as the ``derivative method".
A drawback of the derivative method -- both in its linear \cite{Rattazzi:2008pe,El-Showk:2014dwa,Paulos:2014vya} or semi-definite \cite{Poland:2011ey,Simmons-Duffin:2015qma} programming incarnations -- is the need to include a large number of operators in the bootstrap equations. This makes any, even limited, analytical understanding of the obtained results quite difficult.

A possible approximation scheme is in fact available: ref.~\cite{Pappadopulo:2012jk} has determined the rate of convergence of the Operator Product Expansion (OPE), on which the bootstrap equations are based. This allows us to extract the maximal error from neglecting operators with dimensions larger than some cutoff $\Delta_*$ in the bootstrap equations and thus to consistently truncate them. 
These truncated bootstrap equations can then be evaluated at different points in the $z$-plane.
This method, which we denote as the ``multipoint method", has been previously advocated by Hogervorst and Rychkov in ref.\cite{Hogervorst:2013sma} but has not yet been numerically implemented. The aim of this note is to provide such an implementation and study the resulting bounds.
It is important to emphasize that the method of ref.\cite{Hogervorst:2013sma} combines what are in principle two independent ideas: i) multipoint bootstrap and ii) truncation of the bootstrap equations. One could study i) without ii), or try  to analyze ii) without i). We will not consider these other possibilities here.

We begin in section 2 with a brief review of the results of refs.~\cite{Pappadopulo:2012jk,Hogervorst:2013sma,Rychkov:2015lca} on the convergence of the OPE. We use generalized free theories as a toy laboratory
to test some of the results obtained in ref.\cite{Pappadopulo:2012jk}. We then generalize the results of ref.\cite{Pappadopulo:2012jk} for CFTs with an $O(n)$ global symmetry.

We write the bootstrap equations and set the stage for our numerical computations in section 3.
Our results are then presented in section 4. For concreteness,  we study bounds on operator dimensions and the central charge in 3D and 4D CFTs, with and without an $O(n)$ global symmetry (with no supersymmetry). For these bounds, extensive results are already available in the literature (see e.g.~refs.~\cite{ElShowk:2012ht,El-Showk:2014dwa,Kos:2013tga,Kos:2014bka,Kos:2015mba,Kos:2016ysd,Poland:2011ey,Poland:2010wg, Rattazzi:2010gj, Rattazzi:2010yc,Caracciolo:2014cxa,Iha:2016ppj,Nakayama:2016knq}).  
In particular, we focus our attention on the regions where the 3D Ising and $O(n)$ vector models have been identified.
We show how the results depend on the number $N$ of points in the $z$-plane at which we evaluate the bootstrap equations and the cut-off $\Delta_*$ on the dimension of operators in the bootstrap equations. 
Using values for the dimension of the operator $\phi$ in $O(n)$ vector models available in the literature 
and a fit extrapolation procedure, we then determine the dimensions of the second-lowest $O(n)$ singlet and symmetric-traceless operators $S^\prime$ and $T^\prime$ for $n=2,3,4$. To our knowledge, these have not been obtained before using bootstrap techniques.
Our results are consistent with those from analytical calculations using the $\epsilon$-expansion \cite{Guida:1998bx,Calabrese:2002bm} with a mild tension with the result of ref.\cite{Calabrese:2002bm} for the dimension of $T^\prime$ in the $O(2)$ model.
We notice from our results that the ``kink" in the bound on the dimension of the lowest scalar (singlet) operator in 3D Ising and $O(n)$ vector models is already visible for relatively small $\Delta_*$, while the minimum in the central-charge bound is very sensitive to $\Delta_*$.
For our numerical implementation, we discretize the spectrum and formulate the bootstrap equations as a linear program which we solve using the optimizer \texttt{CPLEX}\footnote{\url{http://www-01.ibm.com/software/commerce/optimization/cplex-optimizer/}} by \texttt{IBM}. Since we focus on the truncated bootstrap equations with relatively low cutoff $\Delta_*$, double precision as used by \texttt{CPLEX} is sufficient for our purposes.
More refined implementations with higher numerical precision, possibly adapting the method and optimizer of refs.\cite{El-Showk:2014dwa,Paulos:2014vya}, are certainly possible.
More details on the numerical implementation are given in section \ref{subsec:technicaldetails}. We conclude in section \ref{conclusions}.


\section{Convergence of the OPE}
\label{OPEconvergence}

We begin with a brief review of the results of refs.~\cite{Pappadopulo:2012jk,Rychkov:2015lca} (see also ref.\cite{Hogervorst:2013sma}) about the convergence of the OPE in a euclidean, reflection positive, CFT in any number of dimensions.\footnote{Bounds on the OPE convergence are obtained in an alternative way using crossing symmetry in ref.\cite{Kim:2015oca}. Interestingly, ref.~\cite{Kim:2015oca} sets bounds which are also valid for finite values of $\Delta_*$ at $z=\bar z=1/2$, though they are relative and not absolute bounds. It would be interesting to explore the approach followed in this paper further. 
We thank Slava Rychkov for having pointed out this reference to us.} For more details see the original references.
Consider the 4-point function of a scalar primary operator $\phi$ with scaling dimension $\Delta_\phi$: 
\be
\label{4-point-function}
\langle \phi(x_1) \phi(x_2) \phi(x_3) \phi(x_4) \rangle \, = \, \frac{g(u,v)}{x_{12}^{2\Delta_\phi}x_{34}^{2\Delta_\phi} } \,,
\ee
where
\be
u \, \equiv \, \frac{x_{12}^2 x_{34}^2}{x_{13}^2 x_{24}^2} \, \quad \text{and} \quad \, v \, \equiv \, \frac{x_{14}^2 x_{23}^2}{x_{13}^2 x_{24}^2} 
\ee
are the conformally-invariant cross-ratios ($x_{ij}\equiv x_i-x_j$).
Applying the OPE to the operator pairs $\phi(x_1) \phi(x_2)$ and $\phi(x_3) \phi(x_4)$ in the 4-point function, one can write
\be
g(u,v) = 1+\sum_{\Delta ,l } \lambda_{\mathcal{O}}^2 \, g_{\Delta,l}(z,\bar z) \,, 
\label{CBExp}
 \ee
where $u = z \bar z$,  $v = (1-z)(1- \bar z)$ and the sum runs over all primary operators $\mathcal{O}$ that appear in the $\phi\times \phi$ OPE with  $\Delta$ and $l$ being respectively their dimension and spin. 
For each primary, the sum over all its descendants is encoded in the conformal block function $g_{\Delta,l}(z,\bar z)$. In a euclidean CFT, $\bar z = z^*$ and the conformal blocks are regular everywhere in the complex $z$-plane, with the exception of a branch-cut along the real  line $[1,+\infty)$.\footnote{The branch-cut is best seen in Lorentzian signature, where $z$ and $\bar z$ are two independent variables. At fixed $\bar z$ (z),  $g_{\Delta,l}(z,\bar z)$ is  a true analytic function in $z$ ($\bar z$) with a branch-cut along the line $[1,+\infty)$.} Thanks to reflection positivity, the OPE coefficients $\lambda_{\mathcal{O}}$ are real and thus $\lambda_{\mathcal{O}}^2 >0$.

Crucial for our considerations will be a bound on the remainder 
\be
\sum_{(\Delta\geq \Delta_*),l} \lambda_{\mathcal{O}}^2 \, g_{\Delta,l}(z,\bar z) 
\label{remainder}
\ee
of the sum in eq.~(\ref{CBExp}) when it is truncated at some primary operator of dimension $\Delta=\Delta_*$.
To determine this bound, one first uses that 
\be 
\left| g_{\Delta,l}(z,\bar z) \right| \, \leq \,   g_{\Delta,l}(|z|,|\bar z|) 
\label{RealBound}
\ee
as follows e.g.~from a representation of the conformal blocks in terms of Gegenbauer polynomials \cite{Hogervorst:2013sma}. 
It is therefore sufficient to estimate the remainder for real $z = \bar z$.
As was found in ref.\cite{Pappadopulo:2012jk}, the most stringent bound is obtained by using the coordinate
\be
\rho(z) = \frac{z}{(1+\sqrt{1-z})^2}\,.
\label{rhoDef}
\ee
The $z$-plane is mapped to the unit disk in $\rho$ and the branch-cut is mapped to the boundary of the disk.
The conformal blocks in $\rho$ are then defined for $|\rho|<1$. 
In the manifestly reflection positive configuration with $\bar \rho= \rho=r$, the function $g(u,v)$ in eq.~(\ref{CBExp}) can be written as\footnote{For simplicity, we use the same symbol to denote the functions $g(u,v)$ and $\tilde{g}(r) = g(u(r),v(r))$ etc. here and below.}
\be
g(r) = 1+\sum_{\Delta ,l} \lambda_{\mathcal{O}}^2 \sum_{n=0}^\infty c_{n}(\Delta,l) r^{\Delta+n}\,,
\label{grRC}
\ee
where $c_n(\Delta,l)$ are positive coefficients whose explicit form is not important here and the sum over $n$ takes into account the contributions from the descendants of each primary.
It is convenient to rewrite $g(r)$ as
\be
g(\beta) = \int_0^\infty \!\! d\Delta \, f(\Delta) e^{-\beta \Delta} \, \quad \text{with} \quad \, f(\Delta) = \sum_{k}\rho_k \, \delta(\Delta-\Delta_k) \,.
\label{grRCInt}
\ee
Here $\beta \equiv - \log r$, $k$ runs over all operators (primaries and their descendants) which are exchanged in the OPE and $f(\Delta)$ is a spectral density with positive coefficients $\rho_k$. 
Again, their explicit form is not relevant for our considerations.

The behaviour of $g(\beta)$ in the limit $\beta\rightarrow 0$ (corresponding to the OPE limit $x_3\rightarrow x_2$, in which case $z \rightarrow r\rightarrow 1$ and $1-z\rightarrow \beta^2/4\rightarrow 0$) is dominated by the exchange of the identity operator and one finds:\footnote{This is true in general only in $d>2$ dimensions. In $d=2$, one has to be careful since 
scalar operators can have arbitrarily small dimensions. See also the discussion after eq.~(\ref{zlim2}).}
\be
g(\beta) \underset{\beta\rightarrow 0}{\sim} 2^{4\Delta_\phi} \beta^{-4\Delta_\phi}\,.
\label{limbeta0}
\ee
Here $a \sim b$ means that $a/b \rightarrow 1$ in the considered limit.
The key observation of ref.\cite{Pappadopulo:2012jk} is that since the coefficients $\rho_k$ are all positive, this asymptotic behaviour determines the leading, large-$\Delta$ behaviour of the integrated spectral density
\be
F(\Delta) = \int_0^\Delta \!\! f(\Delta') \,d\Delta'
\ee
by means of the Hardy-Littlewood tauberian theorem (see e.g.~\cite{tauberian}):\footnote{It is in fact sufficient that the coefficients are all positive for operators with dimension larger than some fixed value $\Delta_0$.\label{PositiveCoefficientsFootnote}}
\be
 F(\Delta) \underset{\Delta\rightarrow \infty}{\sim} \frac{(2 \Delta)^{4\Delta_\phi}}{\Gamma(4\Delta_\phi+1)}\,.
\label{Fdelta}
\ee
The remainder (\ref{remainder}) can then be bounded as follows: We first note that
\be
\sum_{(\Delta\geq \Delta_*),l} \lambda_{\mathcal{O}}^2 \, g_{\Delta,l}(\beta) \leq  \int_{\Delta_*}^\infty \!\! f(\Delta) e^{-\beta \Delta} \,d\Delta\ \, ,
\label{estimate}
\ee
since the r.h.s.~contains contributions from all operators with dimension larger than $\Delta_*$, whereas on the l.h.s.~only primaries with dimension larger than $\Delta_*$ and their descendents contribute.
Using eq.~\eqref{Fdelta}, the r.h.s.~can in turn be bounded as
\bea
 \int_{\Delta_*}^\infty \!\! f(\Delta) e^{-\beta \Delta}  \,d\Delta & \, = \, & \beta  \int_{\Delta_*}^\infty  \!\! e^{-\beta \Delta} (F(\Delta)  - F(\Delta_*)) \,d\Delta \, \leq \, \beta  \int_{\Delta_*}^\infty  \!\! e^{-\beta \Delta} F(\Delta) d\Delta\nn \\
 & \, \simeq \, & \beta  \int_{\Delta_*}^\infty  \!\! e^{-\beta \Delta} \frac{(2 \Delta)^{4\Delta_\phi}}{\Gamma(4\Delta_\phi+1)}d\Delta \, = \, \frac{\beta^{-4\Delta_\phi} \, 2^{4\Delta_\phi}}{\Gamma(4\Delta_\phi+1)} \, \Gamma(4\Delta_\phi+1, \Delta_* \beta) \,,
\label{finalest1}
\eea
where $\Gamma(a,b)$ is the incomplete Gamma function. Clearly, this bound applies for parametrically large values of $\Delta_*$, where eq.~(\ref{Fdelta}) holds.
Using eq.~\eqref{RealBound}, we finally get the bound on the remainder
\be
\Big|\sum_{(\Delta\geq \Delta_*),l} \lambda_{\mathcal{O}}^2 \, g_{\Delta,l}(z,\bar z)\Big| \, \leq \, \frac{(-\log|\rho(z)|)^{-4\Delta_\phi}2^{4\Delta_\phi}}{\Gamma(4\Delta_\phi+1)} \Gamma(4\Delta_\phi+1, - \Delta_* \log |\rho(z)|) \, .
\label{finalest2}
\ee
This is valid in any number $d>2$ of dimensions for 4-point functions with identical scalars.

It was pointed out in ref.~\cite{Rychkov:2015lca} that the conditions for the applicability of the Hardy-Littlewood tauberian theorem in both 3 and 4 dimensions are also fulfilled for the rescaled conformal blocks
\be
\tilde g_{\Delta,l}(r) \equiv (1-r^2)^\gamma g_{\Delta,l}(r)
\ee
with $\gamma=1$. Repeating the derivation reviewed above for a remainder involving the rescaled conformal blocks, it is straightforward to get the alternative bound
\be
\Big|\sum_{(\Delta\geq \Delta_*),l} \lambda_{\mathcal{O}}^2 \, g_{\Delta,l}(z,\bar z)\Big| \, \leq \, \mathcal{R}(z,\bar z,\Delta_*, \Delta_\phi,\gamma)
\label{finalest4}
\ee
with
\be
\mathcal{R}(z,\bar z,\Delta_*, \Delta_\phi,\gamma) \,\equiv \, \frac{(-\log|\rho(z)|)^{-4\Delta_\phi+\gamma} \, 2^{4\Delta_\phi+\gamma}}{\Gamma(4\Delta_\phi+1-\gamma)} \frac{\Gamma(4\Delta_\phi+1-\gamma, - \Delta_* \log |\rho(z)|)}{{(1-|\rho(z)|^2)^\gamma}}\,.
\label{finalest5a}
\ee
For $-\Delta_* \log|\rho(z)|\gg 1$, eq.~(\ref{finalest5a}) can be approximated as
\be
\mathcal{R}(z,\bar z,\Delta_*, \Delta_\phi,\gamma) \, \approx \, \frac{2^{4\Delta_\phi+\gamma} \, \Delta_*^{4\Delta_\phi-\gamma}}{\Gamma(4\Delta_\phi+1-\gamma)} \frac{|\rho(z)|^{\Delta_*}}{(1-|\rho(z)|^2)^\gamma} \,.
\label{finalest5}
\ee
We see that for $|\rho(z)|$ not too close to 1 and $\Delta_* \gtrsim 8 \Delta_\phi$, the bound is more stringent for $\gamma=1$ than for $\gamma=0$. It was furthermore shown in ref.~\cite{Rychkov:2015lca} that in $d=3$ dimensions, $\gamma=1$ is the maximal allowed value such that the Hardy-Littlewood tauberian theorem remains applicable, whereas it was conjectured without proof that the maximal allowed value in $d=4$ dimensions is $\gamma=3/2$. 
Correspondingly we use eq.~\eqref{finalest5a} with $\gamma=1$ for the remainder both in 3 and 4 dimensions in our numerical implementation.\footnote{The fact that eq.~\eqref{finalest4} with $\gamma=0$ is not optimal can be traced to using the inequality \eqref{estimate} in the derivation. In order to make the bound more stringent, one could then alternatively use the series representation in ref.\cite{Hogervorst:2013sma} which includes contributions from primary operators and their descendants separately. Using this series truncated at contributions corresponding to dimension $\Delta_*$ instead of the full conformal blocks $g_{\Delta,l}$ would make the r.h.s.~of the inequality \eqref{estimate} the actual remainder to be bounded. This would thus make eq.~\eqref{finalest4} with $\gamma=0$ more stringent. Here, however, we choose not to follow this approach. The reason is that the representations for the full conformal blocks $g_{\Delta,l}$ can be considerably faster calculated than (our implementation of) the truncated series representation of ref.\cite{Hogervorst:2013sma}.\label{BetterBoundReason}}

The above derivations were based on the existence of a configuration for which the function $g(u,v)$ turns into a positive definite function of a single variable. The remainder is then estimated using the Hardy-Littlewood tauberian theorem. One cannot naively apply these arguments to arbitrary derivatives of $g(u,v)$ w.r.t.~$u$ and $v$, unless the resulting functions remain positive definite and derivatives can be brought inside the absolute value in the l.h.s.~of eq.~(\ref{finalest4}). See the appendix of ref.\cite{El-Showk:2016mxr} for a recent discussion 
on how to estimate the remainder on derivatives of $g(u,v)$. It would be interesting to verify if this allows us to also study truncated bootstrap equations with the derivative method.

\subsection{Comparison with Generalized Free Theories and Asymptotics for $z\rightarrow 1$}

\begin{figure}[!t]
\begin{center}
	\hspace*{0cm} 
	\includegraphics[width=90mm]{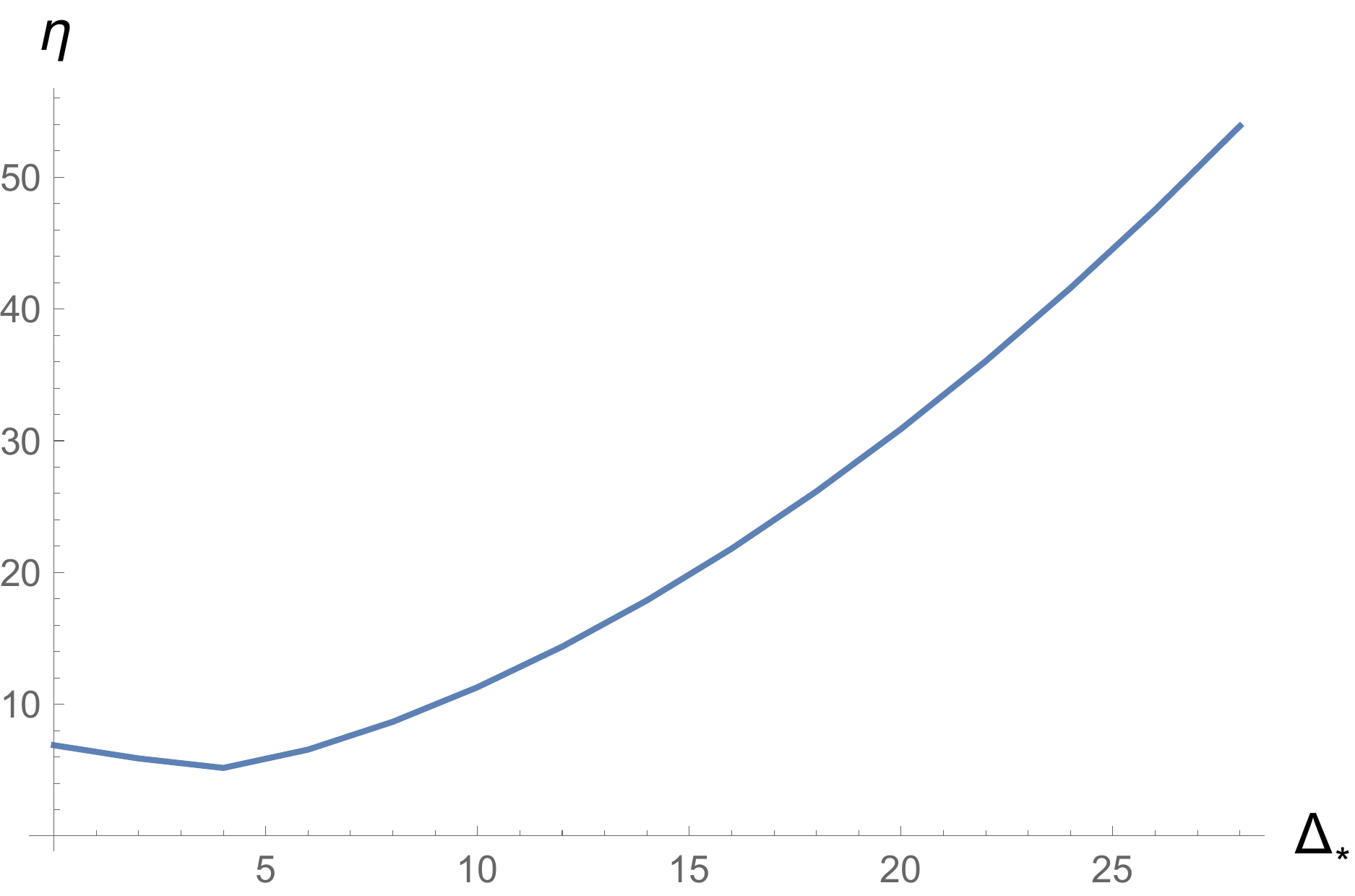}
\end{center}
\caption{\label{fig:1} 
\small
$\eta$ defined in eq.~(\ref{etas}) as a function of $\Delta_*$ in a generalized free theory in $d=4$ dimensions evaluated at the symmetric point $z=\bar z = 1/2$. We have taken $\Delta_\phi=1.5$ and $\gamma=1$.}
\end{figure}

The results reviewed in the previous subsection are based on eq.~(\ref{Fdelta}) which holds in the limit $\Delta_*\rightarrow \infty$. Of course, for any practical use, we need to know the value of $\Delta_*$ beyond which 
we can trust eq.~(\ref{Fdelta}) and thus the bound eq.~(\ref{finalest4}). It is difficult to determine this value for a generic CFT. But we can get useful insights
by considering exactly calculable CFTs, like generalized free theories (sometimes called mean field theories) for which the CFT data are known and the function $g(u,v)$ in eq.~(\ref{4-point-function}) in any number of dimensions reads
\be
g(u,v) \, = \, 1+ u^{\Delta_\phi} + \Big(\frac{u}{v}\Big)^{\Delta_\phi} = 1+ |z|^{2\Delta_\phi} + \Big(\frac{|z|}{|1-z|}\Big)^{2\Delta_\phi}\,.
\label{guvGFT}
\ee
For values of $\Delta_*$  such that eq.~(\ref{Fdelta}) is no good approximation, the r.h.s.~of eq.~(\ref{finalest4}) can clearly still overestimate the actual remainder, leading to no 
inconsistency. On the other hand, if it underestimates the actual remainder, eq.~(\ref{finalest4}) is simply wrong.
We define
\be
\eta   
\, \equiv  \, \frac{\mathcal{R}(z,\bar z,\Delta_*, \Delta_\phi,\gamma)}{\Big|\sum_{(\Delta\geq \Delta_*),l} \lambda_{\mathcal{O}}^2 \, g_{\Delta,l}(z,\bar z)\Big|} 
\label{etas}
\ee
and check if and when $\eta$ 
is smaller than 1, in which case eq.~(\ref{finalest4}) is violated.
The denominator in eq.~(\ref{etas}) is computed as
\be
\sum_{(\Delta\geq \Delta_*),l} \lambda_{\mathcal{O}}^2 \, g_{\Delta,l}(z,\bar z) = g(u,v)-1-\sum_{(\Delta< \Delta_*),l} \lambda_{\mathcal{O}}^2 \, g_{\Delta,l}(z,\bar z)\,.
\ee
In fig.~\ref{fig:1}, we show $\eta$ as a function of $\Delta_*$ evaluated at the symmetric point $z=\bar z = 1/2$.
Notice that at the point of best convergence the actual remainder is always significantly smaller than $\mathcal{R}$, and that the ratio gets bigger and bigger as $\Delta_*$ increases for large $\Delta_*$.
In particular, $\eta$ is greater than 1 for {\it any value} of $\Delta_*$.
We have performed comparisons with GFTs in $d=3$ dimensions with $\gamma=0,1$ and $d=4$ dimensions with $\gamma=0,3/2$ for different values of $z$ and $\Delta_\phi$ within the unitary bounds,
finding analogous qualitative results. 
Somehow unexpectedly, we find that the bound (\ref{finalest4}) is never violated in GFTs, for any value of $\Delta_*$.

When $z\rightarrow 1$, both the numerator and the denominator of $\eta$ in eq.~(\ref{etas}) blow up, since the OPE is not convergent at $z=\bar z=1$. Operators with high scaling dimension are no longer suppressed and the remainder completely dominates the OPE.\footnote{In this limit, the name remainder should actually be used for the finite sum of operators up to $\Delta_*$.}  More precisely, we have
\be
\mathcal{R}(z,\bar z,\Delta_*, \Delta_\phi,\gamma) \, \underset{z,\bar z \rightarrow 1^{-}}{\sim} \, 2^{4\Delta_\phi} (-\log |\rho(z)|)^{-4\Delta_\phi}\,,
\label{zlim1}
\ee
independently of $\gamma$.  Notice that this limit is universal for any CFT that includes in its spectrum a scalar operator with dimension $\Delta_
\phi$, because $z=\bar z \rightarrow 1$ selects the universal identity contribution in the t-channel. This class of CFTs always includes 
a GFT for the operator $\phi$ itself.  In this case the universal nature of the limit is trivially checked using eq.~(\ref{guvGFT}):
\be
g(u,v) \, \underset{z,\bar z \rightarrow 1^{-}}{\sim} \, \frac{1}{|1-z|^{2\Delta_\phi}} \, \underset{z,\bar z \rightarrow 1^{-}}{\sim} \, 2^{4\Delta_\phi} (-\log |\rho(z)|)^{-4\Delta_\phi}\,,
\label{zlim2}
\ee
where in the last equality we have used that $|1-z| \rightarrow (\log|\rho(z)|)^2/4$ in the limit. 

It was found in refs.\cite{Fitzpatrick:2012yx,Komargodski:2012ek} that the spectrum of any Lorentzian CFT resembles that of a GFT for parametrically large spin operators.
In particular, in ref.\cite{Fitzpatrick:2012yx} this has been established by analyzing crossing symmetry in the limit $z\rightarrow 0$ and $\bar z$ fixed for $d>2$, where large twist operators are suppressed. The two-dimensional case is more subtle, because there is no longer a gap between the
identity (which has the minimum twist zero) and the other operators. Indeed, the results of refs.~\cite{Fitzpatrick:2012yx,Komargodski:2012ek} and those of ref.~\cite{Pappadopulo:2012jk} in the euclidean do not  straightforwardly apply for $d=2$.

In the euclidean, operators of any twist should be considered. However, given the results of refs.\cite{Fitzpatrick:2012yx,Komargodski:2012ek},
it is natural to expect that the leading behaviour (\ref{zlim1}) is expected to come from operators with parametrically high dimension and high spin for any CFT,
asymptotically approaching the GFT spectrum in this regime.
It would be interesting to understand within euclidean CFTs, where the twist does not play an obvious role, why this is so.

\subsection{Remainder for CFTs with $O(n)$ Symmetry}
\label{subsec:ONerror}

The generalization of the OPE convergence estimate to CFTs with $O(n)$ global symmetry is straightforward. For concreteness, let us consider scalars $\phi_i$ in the fundamental representation of $O(n)$.
The only non-trivial point is to identify a proper linear combination of 4-point functions 
\be
\langle \phi_i(x_1) \phi_j(x_2) \phi_k(x_3) \phi_l(x_4) \rangle
\ee
that leads to a positive definite series expansion, otherwise the Hardy-Littlewood tauberian theorem does not apply.
A possible choice is 
\be
A_\eta \, \equiv \, \langle \phi_1 \phi_1\phi_1\phi_1 \rangle \, + \, |\eta|^2  \langle \phi_2 \phi_2\phi_2\phi_2 \rangle \, + \, \eta  \langle \phi_1 \phi_1\phi_2\phi_2 \rangle \, + \, \eta^* \langle \phi_2 \phi_2\phi_1\phi_1 \rangle
\, = \, \frac{a_\eta(u,v)}{x_{12}^{2\Delta_\phi}x_{34}^{2\Delta_\phi}}
\,,
\ee
where for simplicity we have omitted the $x$-dependence of the fields. The parameter $\eta$ can in general take an arbitrary complex value, but it is enough for our purposes to consider $\eta=\pm 1$.  For $\bar\rho=\rho=r$ and any $\eta$, this correlator is manifestly positive definite, because it corresponds to the norm of the state
\be
\phi_1 |\phi_1\rangle +\eta \phi_2 |\phi_2\rangle \,. 
\ee
The leading term in $a_\eta(u,v)$ for $x_2\rightarrow x_3$ is given by the exchange of the identity operator in the first two correlators and hence is independent of $\eta$.
On the other hand, expanding in conformal blocks in the (12)-(34) channel, we have \cite{Rattazzi:2010yc}
\be
A_\eta = \frac{1}{x_{12}^{2\Delta_\phi}x_{34}^{2\Delta_\phi}}\Big( 2(1+\eta)  \Big( 1 +\sum_{S^+} \lambda_S^2 \, g_{\Delta,l}(u,v)\Big)  + 4 \Big(1-\frac{1+\eta}{n}\Big)  \sum_{T^+} \lambda_T^2 \, g_{\Delta,l}(u,v) \Big)\,,
\label{Aeta}
\ee
where $S$ and $T$ denote operators in the singlet and rank-two symmetric representations of $O(n)$, respectively. Both sums run over even spins.
We can now repeat essentially verbatim the derivation below eq.~(\ref{rhoDef}).
For $\eta = -1$, this gives rise to the bound
\be
\Bigl| \sum_{(\Delta \geq \Delta_*),l} \lambda_T^2 \, g_{\Delta,l}(z,\bar z) \Bigr| \, \leq \, \frac 12 \mathcal{R}(z,\bar z,\Delta_*, \Delta_\phi,\gamma)
 \, ,
\label{EstimateT}
\ee
where $\mathcal{R}$ is given in eq.~(\ref{finalest5a}). The factor 1/2 with respect to the non-symmetric case arises because the identity operator is exchanged in two correlators but a factor 4 is present in the  second term in the r.h.s.~of eq.~(\ref{Aeta}). 
For $\eta=1$ we similarly get
\be
\Bigl|  \sum_{(\Delta \geq \Delta_*),l} \left( \lambda_S^2 \, g_{\Delta,l}(z,\bar z) +\big(1-\frac 2n\big) \lambda_T^2 \, g_{\Delta,l}(z,\bar z)\right) \Bigr| \leq \frac 12 \mathcal{R}(z,\bar z,\Delta_*, \Delta_\phi,\gamma)\,.
\label{EstimateTS}
\ee

Another positive definite linear combination of correlators is 
\be
B_\eta \equiv \langle \phi_2 \phi_1\phi_1\phi_2 \rangle+ |\eta|^2 \langle \phi_1 \phi_2\phi_2\phi_1 \rangle+ \eta \langle \phi_2 \phi_1\phi_2\phi_1 \rangle+\eta^*  
  \langle \phi_1 \phi_2\phi_1\phi_2 \rangle =\frac{b_\eta(u,v)}{x_{12}^{2\Delta_\phi}x_{34}^{2\Delta_\phi}}
\,,
\ee
corresponding to the norm of the state
\be
\phi_1 |\phi_2\rangle +\eta \phi_2 |\phi_1\rangle \,. 
\ee
Again, we consider $\eta=\pm 1$. 
In the (12)-(34) channel the correlator $B_\eta$  can be written as\footnote{In our normalization conventions for the conformal blocks, the squared OPE coefficients $\lambda_{S,T,A}^2$ are all positive.}
\be
B_\eta = \frac{1}{x_{12}^{2\Delta_\phi}x_{34}^{2\Delta_\phi}}\Big( 2(1+\eta)  \sum_{T^+} \lambda_T^2 \, g_{\Delta,l}(u,v) + 2(1-\eta)  \sum_{A^-} \lambda_A^2 \, g_{\Delta,l}(u,v)  \Big)\,,
\label{Beta}
\ee
where $A$ stands for operators in the rank-two antisymmetric representation of $O(n)$. The first sum runs over even spins, whereas for the second one they are odd. As before, 
the leading term in $b_\eta(u,v)$ for $x_2\rightarrow x_3$ is given by the exchange of the identity operator in the first two correlators and is independent of $\eta$.
For $\eta = 1$, eq.~(\ref{Beta}) gives rise to the same bound given in eq.~(\ref{EstimateT}), while for $\eta=-1$ we have
\be
\Bigl| \sum_{(\Delta \geq \Delta_*),l} \lambda_A^2 \, g_{\Delta,l}(z,\bar z) \Bigr| \, \leq \, \frac 12 \mathcal{R}(z,\bar z,\Delta_*, \Delta_\phi,\gamma) \,.
\label{EstimateA}
\ee

It is straightforward to see that the bounds (\ref{EstimateT}), (\ref{EstimateTS}) and (\ref{EstimateA}) are the best that can be obtained.
Indeed, in the free-theory limit one has $\lambda_S^2 = \lambda^2/n$, $\lambda_T^2 = \lambda_A^2= \lambda^2/2$ with $\lambda^2$ being the OPE coefficients for a single free field 
(see e.g.~eq.~(5.11) in ref.~\cite{Caracciolo:2014cxa}). The above three bounds then reduce to eq.~(\ref{finalest4})
which is known to give the best bound on the r.h.s.~of eq.~\eqref{estimate} (see however footnote \ref{BetterBoundReason}) \cite{Pappadopulo:2012jk}.
Any potentially better bound for $O(n)$ theories should in particular apply to the free theory, but would then be in contradiction with the results of ref.~\cite{Pappadopulo:2012jk}.

The above bounds will be used in the next section to bound the remainder of the bootstrap equations in CFTs with an $O(n)$ global symmetry.

\section{Bootstrapping with Multiple Points}

The bootstrap equation for a 4-point function with identical scalars $\phi$ with scaling dimension $\Delta_\phi$ in any number of dimensions is given  by the sum rule (see refs.\cite{Rychkov:2016iqz,Simmons-Duffin:2016gjk} for pedagogical reviews)
\begin{align}
\label{bootstrapequation}
\sum_{\Delta ,l } \lambda_{\mathcal{O}}^2 \, \mathcal{F}_{\Delta_\phi,\Delta,l}(z, \bar z) \, = \, u^{\Delta_\phi} - v^{\Delta_\phi} \, , \hspace{.8cm} \mathcal{F}_{\Delta_\phi,\Delta,l}(z, \bar z) \, \equiv \, v^{\Delta_\phi} g_{\Delta,l}(u,v)- u^{\Delta_\phi} g_{\Delta,l}(v,u) \,.
\end{align} 
Splitting the sum into two parts, for dimensions smaller and larger than a cutoff $\Delta_*$, we can write
\be
\label{bootstrapequation-approx}
\sum_{(\Delta < \Delta_*),l } \lambda_{\mathcal{O}}^2 \, \mathcal{F}_{\Delta_\phi,\Delta,l}(z, \bar z) \, = \, u^{\Delta_\phi} - v^{\Delta_\phi} \,+ \, \mathcal{E}(z,\bar z,\Delta_*, \Delta_\phi) \,.
\ee
Using eq.~\eqref{finalest4}, the remainder $\mathcal{E}$ of the sum rule is bounded by
\be 
\label{ErrorEstimate}
\left| \mathcal{E}(z,\bar z) \right| \, \leq \,\mathcal{E}_{\rm max}(z,\bar z) \, \equiv \, v^{\Delta_\phi} \, \mathcal{R}(z,\bar z) \, + \, u^{\Delta_\phi} \, \mathcal{R}(1-z,1- \bar z) \, ,
\ee
where we have omitted the dependence on $\Delta_*$, $\Delta_\phi$ and $\gamma$.
The truncated sum rule \eqref{bootstrapequation-approx} still involves a generally unknown  spectrum of operators up to dimension $\Delta_*$. In order to make it amenable to numerical analysis, we discretize the spectrum and make the ansatz\footnote{Alternatively, one could adapt the approach of ref.~\cite{El-Showk:2014dwa} to the multipoint method.}
\be 
\label{lDeltavec}
\Big\{  (0,\frac{d-2}{2}) \,,
(0,\frac{d-2}{2}+\Delta_{\rm step})\,, 
\ldots \,,
(0,\Delta_*)\,,
(2,d) \,,
(2,d+\Delta_{\rm step}) \,, 
(2,\Delta_*)\,,
\ldots \,,
(l_{\rm max},\Delta_*)\Big\} 
\ee
for the quantum numbers (spin,dimension) of the operators that can appear in the truncated sum rule. For each spin $l$, the dimension runs in steps of size $\Delta_{\rm step}$ from the unitarity bound $\Delta_{\rm min}^{d,l} \equiv l + (d-2)/(1+\delta_{l0})$ to the cutoff $\Delta_*$ (or a value close to that, depending on $\Delta_{\rm step}$). Accordingly, $l_{\rm max}$ is the largest spin for which the unitarity bound is still below the cutoff, $\Delta_{\rm min}^{d,l_{\rm max}}< \Delta_*$. In practice, we vary the step size $\Delta_{\rm step}$ somewhat depending on the spin and dimension. This is discussed in more detail in sec.~\ref{subsec:technicaldetails}. We find that the bounds converge when going to smaller $\Delta_{\rm step}$, meaning that the discretization does not introduce any artifacts into our calculation.

We similarly choose a finite number of points $z_i$ in the $z$-plane where the sum rule is evaluated. The details of our choice for this distribution of points are discussed in sec.~\ref{sec:pointchoice}. 
Together with the discretization of operator dimensions, this turns eq.~\eqref{bootstrapequation-approx} into the matrix equation
\be 
\label{DiscretizedBootstrapEquation}
\mathcal{M} \cdot  \vec{\rho} \, = \, \vec{\sigma} \, + \, \vec{\epsilon}  \, .
\ee
The elements of the matrix $\mathcal{M}$ are the functions $\mathcal{F}_{\Delta_\phi,\Delta,l}(z, \bar z)$ evaluated for the different quantum numbers in eq.~\eqref{lDeltavec} along the rows and for the different points $z_i$ along the columns. Furthermore, the vector $\vec{\rho}$ consists of the squared OPE coefficients $\lambda_{\mathcal{O}}^2$ of the operators corresponding to the quantum numbers in eq.~\eqref{lDeltavec} and
\be 
\vec{\sigma} \, \equiv \, \left( \begin{matrix}
                      |z_1|^{2\Delta_\phi} \, - \, |1-z_1|^{2 \Delta_\phi}\\  |z_2|^{2\Delta_\phi} \, - \, |1-z_2|^{2 \Delta_\phi} \\ \vdots
                     \end{matrix} \right) 
                     \quad \, \text{and} \, \quad 
                      \vec{\epsilon} \, \equiv \, 
                          \left( \begin{matrix}
                                  \mathcal{E}(z_1,\bar z_1,\Delta_*,\Delta_\phi) \\
                                  \mathcal{E}(z_2,\bar z_2,\Delta_*,\Delta_\phi) \\
                                  \vdots
                                 \end{matrix} \right)  \, .
\ee
Using the bound \eqref{ErrorEstimate}, we then obtain the matrix inequality
\be 
\label{LinearProgram}
 \left( \begin{matrix}
                      \mathcal{M} \\ -\mathcal{M} 
                     \end{matrix} \right)  \,\vec{\rho} \,  \geq \, \left( \begin{matrix}
                      \vec{\sigma} - \vec{\epsilon}_{\rm max} \\  -\vec{\sigma} - \vec{\epsilon}_{\rm max} 
                     \end{matrix} \right) \, ,
\ee
where $\vec{\epsilon}_{\rm max}$ is defined as $\vec{\epsilon}$ but with $\mathcal{E}$ replaced by $\mathcal{E}_{\rm max}$. This is the starting point for our numerical calculations. In order to determine bounds on OPE coefficients, we search for vectors $\vec{\rho}$ which satisfy eq.~\eqref{LinearProgram} and extremize the entry corresponding to that OPE coefficient. For bounds on the dimension of the lowest-lying scalar operator, on the other hand, we make an assumption on this dimension and drop all scalar operators with smaller dimension from our ansatz \eqref{lDeltavec}. This gap then allows for a consistent CFT only if there exists a vector $\vec{\rho}$ which satisfies eq.~\eqref{LinearProgram} with the reduced ansatz. By trying different assumptions, we can determine the maximal allowed gap. Both problems are linear programs which can be solved using fast numerical routines. An advantage of solving eq.~\eqref{LinearProgram} is that the vector $\vec{\rho}$ gives us the spectrum of operators and their OPE coefficients of a potential CFT living at the boundary of the allowed region. This has been used before in ref.\cite{El-Showk:2014dwa}.\footnote{The data of CFTs at the boundary of the allowed region can also be obtained from the `dual' method originally developed in ref.~\cite{Rattazzi:2008pe} by using the extremal functional method of ref.~\cite{ElShowk:2012hu}.}  

We also consider CFTs with an $O(n)$ global symmetry. For an external scalar operator in the fundamental representation of $O(n)$, the sum rule reads \cite{Rattazzi:2010yc}
\be
\sum_{S^+} \lambda_{S}^2 \left(\begin{array}{c}
0\\
\mathcal{F} \\
\mathcal{H} \end{array}\right)
+\sum_{T^+} \lambda_{T}^2 \left(\begin{array}{c}
\mathcal{F} \\
(1-\frac 2n)\mathcal{F} \\
-(1+\frac 2n) \mathcal{H} \end{array}\right)
+\sum_{A^-} \lambda_{A}^2 \left(\begin{array}{c}
-\mathcal{F} \\
\mathcal{F} \\
-\mathcal{H} \end{array}\right) =
 \left(\begin{array}{c}
0\\
u^{\Delta_\phi} -v^{\Delta_\phi} \\
-u^{\Delta_\phi} -v^{\Delta_\phi} \end{array}\right) \, ,
\label{CrossingSON}
\ee 
where 
$ \mathcal{H}_{\Delta_\phi,\Delta,l}(z, \bar z) \, \equiv \, v^{\Delta_\phi} g_{\Delta,l}(u,v)+ u^{\Delta_\phi} g_{\Delta,l}(v,u)$
and we have suppressed the arguments of the functions $\mathcal{F}$ and $\mathcal{H}$.
Splitting the sums in eq.~\eqref{CrossingSON} into two parts, for dimensions smaller and larger than a cutoff $\Delta_*$, we can write
\be
\sum_{\substack{S^+ \\ \Delta < \Delta_*}} \lambda_{S}^2 \left(\begin{array}{c}
0\\
\mathcal{F} \\
\mathcal{H} \end{array}\right)
+\sum_{\substack{T^+ \\ \Delta < \Delta_*}} \lambda_{T}^2 \left(\begin{array}{c}
\mathcal{F} \\
(1-\frac 2n)\mathcal{F} \\
-(1+\frac 2n) \mathcal{H} \end{array}\right)
+\sum_{\substack{A^- \\ \Delta < \Delta_*}} \lambda_{A}^2 \left(\begin{array}{c}
-\mathcal{F} \\
\mathcal{F} \\
-\mathcal{H} \end{array}\right) =
 \left(\begin{array}{c}
\mathcal{E}_1\\
u^{\Delta_\phi} -v^{\Delta_\phi} + \mathcal{E}_2 \\
-u^{\Delta_\phi} -v^{\Delta_\phi} + \mathcal{E}_3 \end{array}\right).
\label{CrossingSONapprox}
\ee 
Using eqs.~(\ref{EstimateT}), (\ref{EstimateTS}) and (\ref{EstimateA}), we obtain the bounds on the remainders
 \be
\left| \mathcal{E}_{1,2}(z,\bar z) \right| \, \leq \,\mathcal{E}_{\rm max}(z,\bar z) \,,\quad  \quad 
\left| \mathcal{E}_{3}(z,\bar z) \right| \, \leq 2 \,\mathcal{E}_{\rm max}(z,\bar z) \,,
\ee
with $\mathcal{E}_{\rm max}$ defined as in eq.~\eqref{ErrorEstimate}. Discretizing the space of operator dimensions as in eq.~\eqref{lDeltavec} and evaluating the sum rule at a finite set of points $z_i$, we again obtain a matrix inequality of the form \eqref{LinearProgram}. This is the starting point for our numerical calculations for CFTs with $O(n)$ global symmetry.

\subsection{Choice of Points}
\label{sec:pointchoice}

\begin{figure}[!t]
\begin{center}
	\hspace*{0cm} 
	\includegraphics[width=100mm]{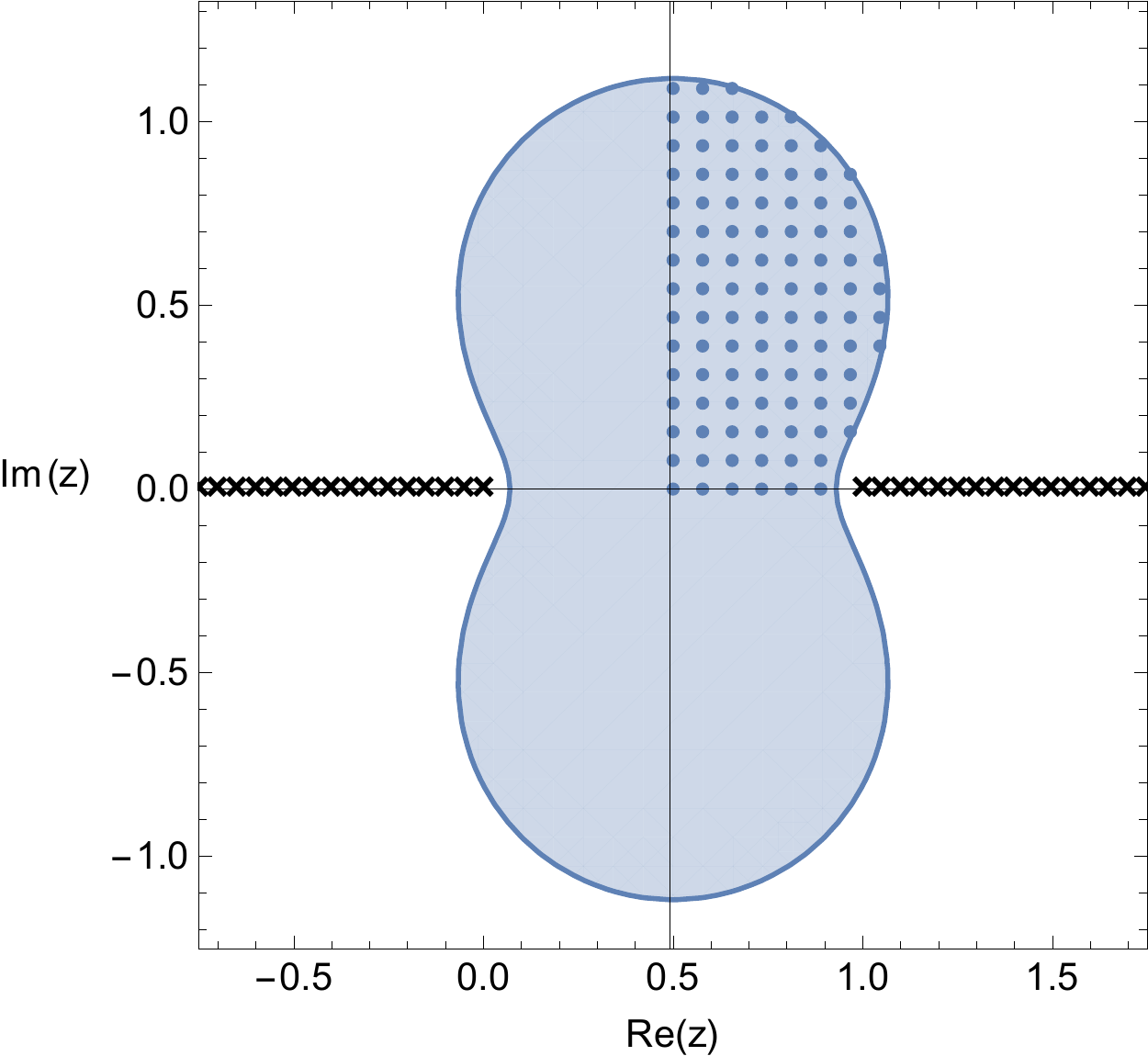}\\
	\end{center}
\vspace*{-0.2cm}
\caption{\label{fig:zplane} 
\small
The region in the $z$-plane with $\lambda(z)\leq 0.6$ and a sample of 100 points in a fundamental domain of that region. The crossed lines are the two branch-cuts where the bootstrap equations do not converge.}
\end{figure}

An important choice for the multipoint method is the distribution of points in the $z$-plane at which the bootstrap equations are evaluated. Using the symmetries $z\leftrightarrow \bar z$ and $z\leftrightarrow (1-z)$, $\bar z\leftrightarrow 1-\bar z$ of the bootstrap equations, we can restrict these points to the region $\operatorname{Re}(z)\geq 1/2$ and $\operatorname{Im}(z)\geq 0$ of the $z$-plane. The remainder of the truncated sum rule is controlled by $|\rho(z)|$ and $|\rho(1-z)|$ (cf.~eqs.~\eqref{finalest5} and \eqref{ErrorEstimate}). Guided by this, we introduce the measure
\be 
\label{measure}
\lambda(z) \, \equiv \,  |\rho(z)| \, + \, |\rho(1-z)| \,,
\ee
and consider points with $\lambda(z)\leq \lambda_c$ for some constant $\lambda_c$.
It is desirable to choose $\lambda_c$ and the distribution of points within that region in such a way that the obtained bounds are as stringent as possible.
We have performed extensive scans over different values for $\lambda_c$ and distributions with different density profiles and have found that a flat profile leads to as good or better bounds than more complicated profiles. We therefore choose the former and put points on a grid centered at $z=1/2$. 
The grid spacing is chosen such that the desired number of points is within the region $\lambda(z)\leq \lambda_c$, $\operatorname{Re}(z)\geq 1/2$ and $\operatorname{Im}(z)\geq 0$. We have then found that 
\be
\lambda_c = 0.6
\ee
gives the best bounds for all cases that we have studied.\footnote{In more detail, we have considered bounds on the central charge and the dimension of the lowest-dimensional scalar operator, in 3D and 4D, with $O(n)$ and without symmetry, and with different choices for the number of points $N$ and the cutoff $\Delta_*$. It is remarkable that $\lambda_c=0.6$ (within $\pm 0.02$, the resolution of our scan) comes out as the optimal choice for such a variety of cases.} In fig.~\ref{fig:zplane}, we show the corresponding region in the $z$-plane and a sample distribution of 100 points. 

In order to test the influence of the choice of measure on the bounds, we have performed further scans with $\lambda(z) \equiv  \max(|\rho(z)|,|\rho(1-z)|)$ proposed in ref.~\cite{Hogervorst:2013sma} and $\lambda(z) \equiv |z-1/2|$ (for the latter we have removed points at or close to the branch-cuts). We have found that, once the optimal $\lambda_c$ is chosen, the bounds obtained with these measures are indistinguishable from those obtained with eq.~\eqref{measure}. This indicates that the precise form of the region within which points are sampled has only a marginal effect on the quality of the bounds.

\section{Results}

We now present the results of our numerical analysis. 
In subsection \ref{subsec:IsingOn}, we study bounds on the dimension of the lowest-dimensional scalar operator in the OPE and bounds on the central charge in 3D CFTs, focusing in particular on the 
regions where the 3D Ising and $O(n)$ models have been identified. In subsection \ref{subsec:4D} we then study the same bounds for generic 
4D CFTs. We analyze in particular  how our results depend on the number $N$ of points chosen in the $z$-plane,
and on the cutoff $\Delta_*$. In subsection \ref{subsec:spectrumON} we give a closer look at the spectrum of the 3D $O(n)$ models
and determine the operator dimensions of the first two scalar operators in the singlet and rank-two symmetric representation of $O(n)$. 

Before presenting our results, it is important to emphasize an important difference between the multipoint and the derivative bootstrap methods.
As mentioned in the introduction, in the latter we do not have a reliable way of truncating the OPE series defining the bootstrap equations at some 
intermediate dimension $\Delta_*$, because we do not have a reliable estimate of the resulting error. We are therefore forced to have $\Delta_*$ as large as possible to minimize this error and can only check a posteriori if the chosen $\Delta_*$ was sufficient.\footnote{We are a bit sloppy here in order to keep the discussion simple and get to the point.
For instance, in numerical methods based on semi-definite programming one is able to include all operator dimensions continuously up to infinity.
The rough analogue of our $\Delta_*$ in that case is the maximum spin of the primary operators entering the OPE which are taken into account for the numerical implementation.}
More than $\Delta_*$ (or its analogue), the key parameter that controls the accuracy of the method is given by the total number of derivatives $N_D$ that are applied to the bootstrap 
equations. Of course, the larger $N_D$ is, the better are the bounds. 
The accuracy is then limited by the largest $N_D$ that allows the calculation to be performed within an acceptable amount of time with the available computing resources.

In the multipoint method, on the other hand, we can reliably vary $\Delta_*$ due to the bound on the remainder of the truncation discussed in sec.~\ref{OPEconvergence}. In addition, we can also vary the number $N$ of points in the $z$-plane which is the analogue of $N_D$ in the derivative method. 
The parameter region for the multipoint method corresponding to the typical bootstrap analysis with the derivative method is then very large $\Delta_*$ and $N$ as large as possible given the available computing resources.
In this paper, on the other hand, we are mostly interested in the regime where $\Delta_*$ is {\it not} very large,  with values $\mathcal{O}(10)$-$\mathcal{O}(20)$.
We find that for this range of $\Delta_*$, the results converge for $N\sim \mathcal{O}(100)$ and do not improve further if $N$ is increased. This corresponds to the fact that the rank of the matrix $\mathcal{M}$ in the discretized bootstrap equation \eqref{DiscretizedBootstrapEquation} is then $\mathcal{O}(100)$.
Note that since \texttt{CPLEX} is limited to double precision, we also cannot take $\Delta_*$ arbitrarily large. Due to the excellent speed of \texttt{CPLEX}, on the other hand, we have found that taking $N$ large enough so that the bounds converge is no limiting factor.

\subsection{3D Ising and $O(n)$ Models}
\label{subsec:IsingOn}

\begin{figure}[!t]
\begin{center}
\includegraphics[width=110mm]{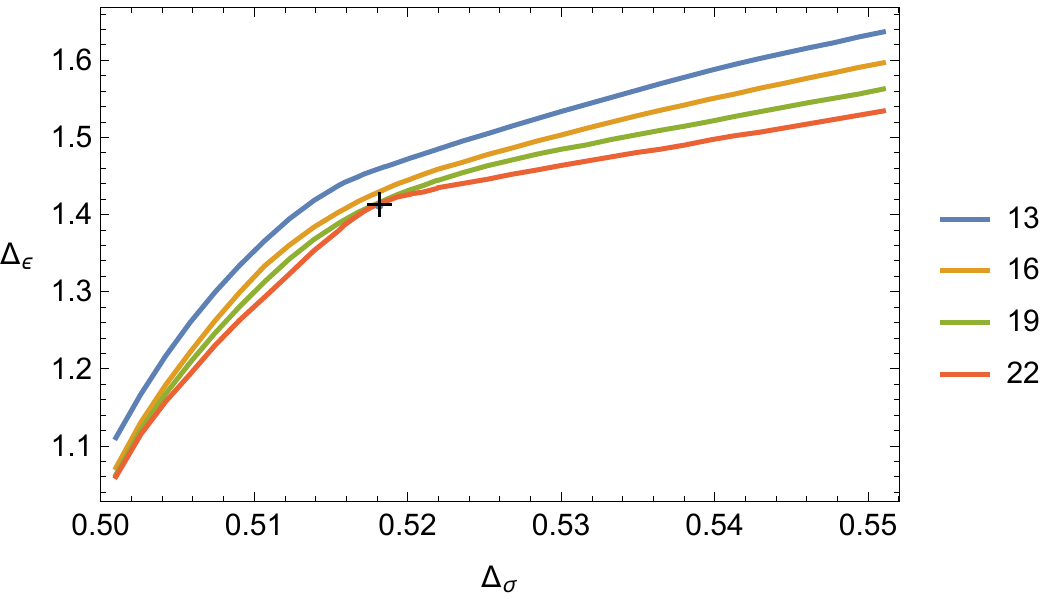}
\end{center}
\vspace*{-0.2cm} 
\caption{\label{fig:2} 
\small
Bounds on $\Delta_\epsilon$ as a function of $\Delta_\sigma$ for $N=100$ points and different values of $\Delta_*$. The regions above the lines are excluded. 
The black cross marks the precise values of $\Delta_\sigma$ and $\Delta_\epsilon$ for the 3D Ising model as determined in ref.\cite{El-Showk:2014dwa}. The curves and the labels in the legend have the same order from top to bottom.}
\begin{center}
	\hspace*{0cm} 
	\includegraphics[width=110mm]{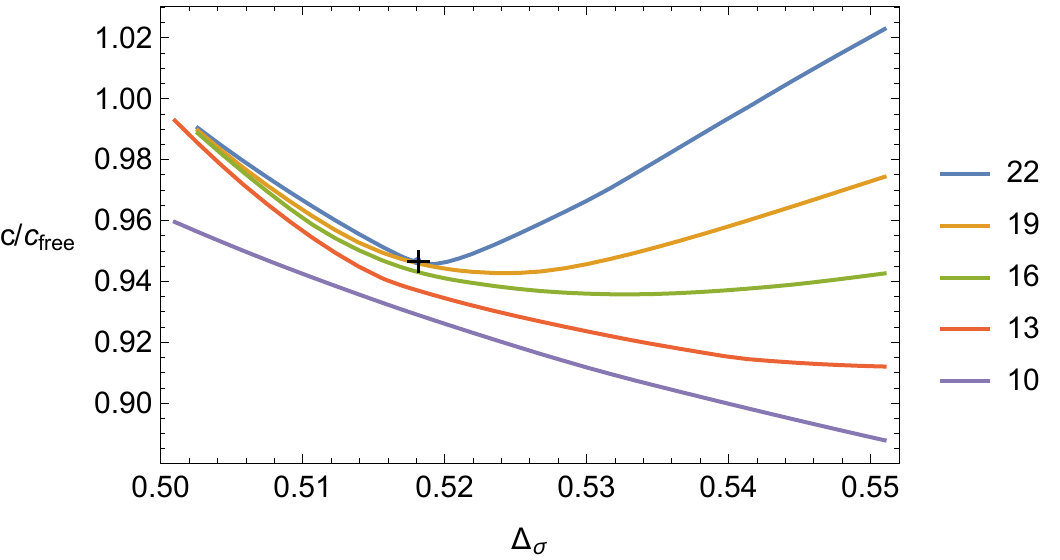}
	\end{center}
\vspace*{-0.2cm}
\caption{\label{fig:3} 
\small
Bounds on the central charge $c$ as a function of $\Delta_\sigma$ for $N=100$ points and different values of $\Delta_*$. A gap $\Delta_\epsilon>1.1$ has been assumed. The regions below the lines are excluded. 
The black cross marks the precise values of $\Delta_\sigma$ and $c$ for the 3D Ising model as determined in ref.\cite{El-Showk:2014dwa}. The curves and the labels in the legend have the same order from top to bottom.}
\end{figure}

The most remarkable numerical results from the conformal bootstrap have been obtained in 3D CFTs.
One interesting bound to study is on the dimension of the lowest-dimensional scalar operator appearing in the OPE. We denote this operator by $\epsilon$ and the operator that is used to derive the bootstrap equations by $\sigma$. It was noted in ref.~\cite{ElShowk:2012ht} that the 3D Ising model sits at a special point, a kink, at the boundary of the allowed region of $\Delta_\epsilon$ as a function of $\Delta_\sigma$.
The Ising model is similarly special with respect to the bound on the central charge $c$ as a function of $\Delta_\sigma$, sitting again at the boundary of the excluded region,
at the point where $c$ is minimized \cite{ElShowk:2012ht,El-Showk:2014dwa}. Note, however, that the theory minimizing $c$ does not actually correspond to the 3D Ising model, but rather to some
exotic theory with $\Delta_\epsilon<1$. Most likely this theory is unphysical (though we are not aware of a solid argument to dismiss it). In practice this theory is removed by
assuming a gap in the operator spectrum such that $\Delta_\epsilon>1$.  Independently of the nature of this theory, the condition  $\Delta_\epsilon>1$  is satisfied by the Ising model
and can be legitimately imposed if we are interested in this particular 3D CFT.

In fig.~\ref{fig:2}, we show the bound on $\Delta_\epsilon$ as a function of $\Delta_\sigma$ for $N=100$ points and different values of $\Delta_*$. Notice how the kink shows up already for $\Delta_*=13$ and converges quite quickly as $\Delta_*$ increases.  In fig.~\ref{fig:3}, we show the bound on the central charge $c$ (normalized to the central charge $c_{\rm free}$ of a free scalar theory) as a function of $\Delta_\sigma$ for $N=100$ points and different values of $\Delta_*$.
The gap $\Delta_\epsilon > 1.1$ is assumed in the operator spectrum.
A lower bound on $c$ is obtained even for $\Delta_*=10$, but the convergence when going to larger $\Delta_*$ is now much slower than for the bound on $\Delta_\epsilon$. A minimum is visible starting from $\Delta_*=16$ but even at $\Delta_*=22$ it is a bit shifted to the right with respect to its actual value. We have still not reached the asymptotic value for $\Delta_*$. Unfortunately, we cannot get reliable results for much higher $\Delta_*$ because the numerical accuracy of \texttt{CPLEX} is limited to double precision. 
Nevertheless, it is clear from comparing figs.~\ref{fig:2} and \ref{fig:3} that the lower bound on $c$ is more ``UV sensitive" than the bound on $\Delta_\epsilon$.
In both figures, the crosses mark the location of the 3D Ising model, as determined in ref.~\cite{El-Showk:2014dwa}.

\begin{figure}[!t]
\begin{center}
	\hspace*{0cm} 
	\includegraphics[width=110mm]{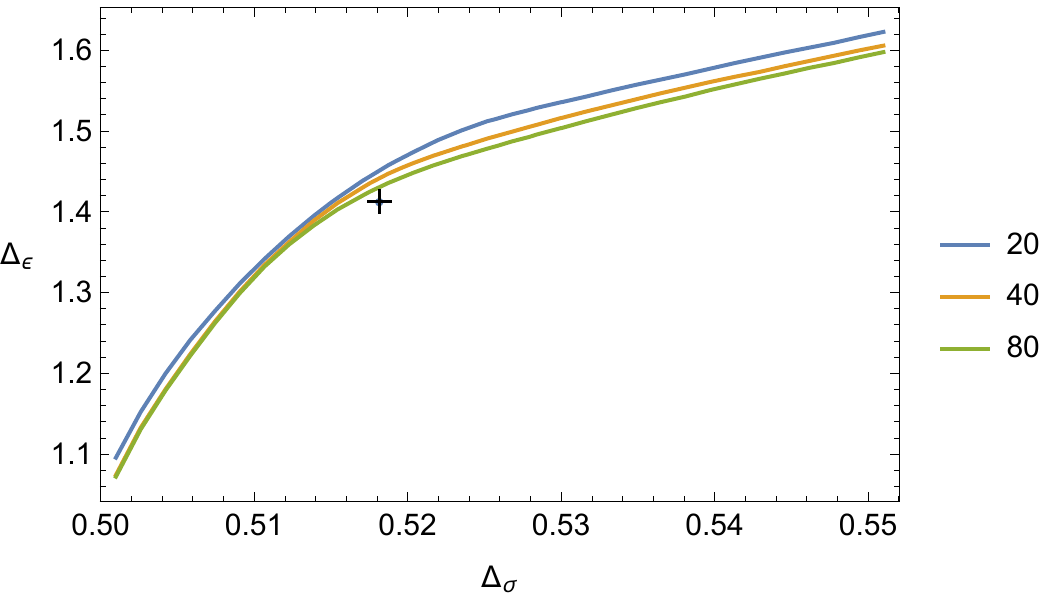}
	\end{center}
\vspace*{-0.2cm}
\caption{\label{fig:4} 
\small
Bounds on  $\Delta_\epsilon$  as a function of $\Delta_\sigma$ for fixed $\Delta_*=16$ and different values of $N$. The regions above the lines are excluded. The black cross marks the precise values of $\Delta_\sigma$ and $\Delta_\epsilon$ for the 3D Ising model as determined in ref.\cite{El-Showk:2014dwa}. The curves and the labels in the legend have the same order from top to bottom.}
\begin{center}
	\hspace*{0cm} 
	\includegraphics[width=110mm]{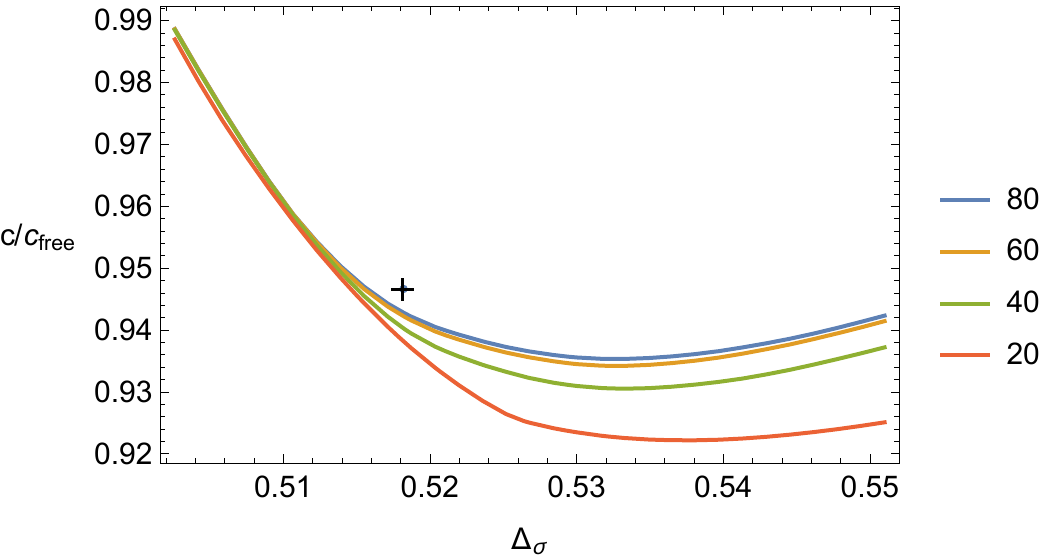}
\end{center}
\vspace*{-0.2cm}
\caption{\label{fig:4a} 
\small
Bounds on the central charge $c$ as a function of $\Delta_\sigma$ for fixed $\Delta_*=16$ and different values of $N$. The gap $\Delta_\epsilon>1.1$ is assumed. The regions below the lines are excluded. The black cross marks the precise values of $\Delta_\sigma$ and $c$ for the 3D Ising model as determined in ref.\cite{El-Showk:2014dwa}. The curves and the labels in the legend have the same order from top to bottom.}
\end{figure}

In order to quantify the dependence of our results on the number $N$ of points,  we show in figs.~\ref{fig:4} and \ref{fig:4a} the bounds on respectively $\Delta_\epsilon$ and $c$ as a function of $\Delta_\sigma$ for different values of $N$ at fixed $\Delta_*=16$. We see that in both cases the convergence in $N$ is quite fast, with $N=40$ for $\Delta_\epsilon$ and $N=60$ for $c$ being already an excellent approximation. Notice that for increasing $N$, the bound on $\Delta_\epsilon$ converges faster than the bound on $c$, similar to the dependence on $\Delta_*$. We have studied the dependence on $N$ also for different values of $\Delta_*$ and have found as expected that the value $N_*$ beyond which no significant improvement in the bounds is observed increases with $\Delta_*$. The dependence is however very mild for the central charge $c$ and barely observable for $\Delta_\epsilon$.
This is still a reflection of the different ``UV  sensitivities" of the two quantities. In all cases, $N_*\lesssim \mathcal{O}(100)$ up to $\Delta_*=24$.

\begin{figure}[!t]
\begin{center}
\hspace*{0cm} 
\includegraphics[width=110mm]{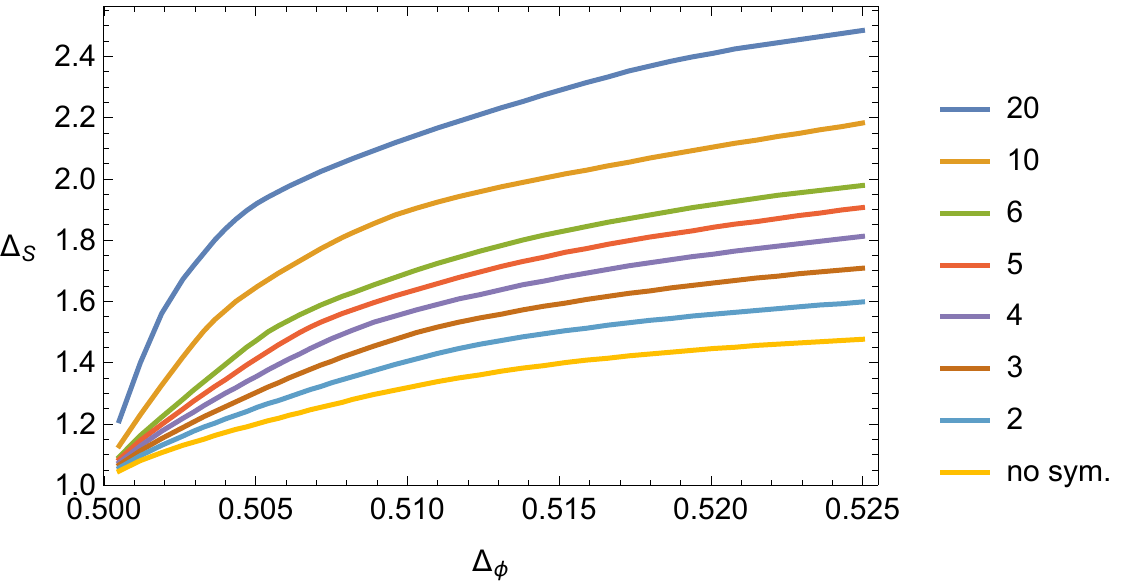}
\end{center}
\vspace*{-0.2cm}
\caption{\label{fig:5} 
\small
Bounds on $\Delta_S$ as a function of $\Delta_\phi$ for 3D CFTs with different $O(n)$ symmetries, with $\phi$ in the fundamental representation of $O(n)$. The regions above the lines are excluded. All the bounds have been determined using $N=80$ points and $\Delta_*=16$. The curves and the labels in the legend have the same order from top to bottom.}
\begin{center}
	\hspace*{0cm} 
	\includegraphics[width=110mm]{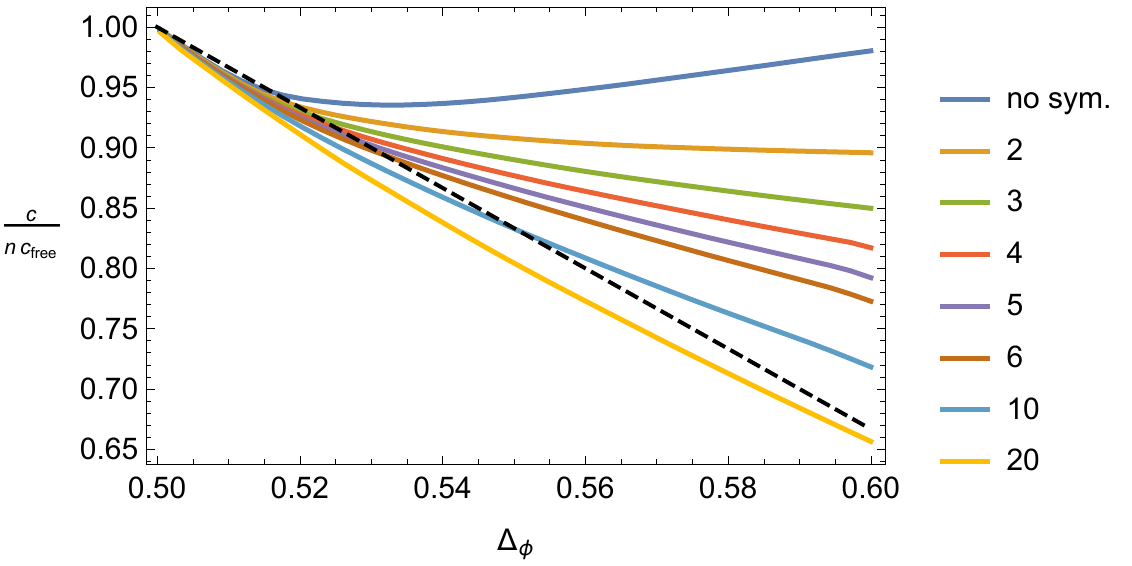}
	\end{center}
\vspace*{-0.2cm}
\caption{\label{fig:6} 
\small
Bounds on the central charge $c$ as a function of $\Delta_\phi$ for 3D CFTs with different $O(n)$ symmetries, with $\phi$ in the fundamental representation of $O(n)$. The regions below the lines are excluded. All the bounds have been determined using $N=80$ points and $\Delta_*=16$ with gaps $\Delta_S>1$ and $\Delta_T>1$ assumed.
The dashed line is the leading large-$n$ prediction. The curves and the labels in the legend have the same order from top to bottom.}
\end{figure}

\begin{figure}[!t]
\begin{center}
\hspace*{0cm} 
\includegraphics[width=110mm]{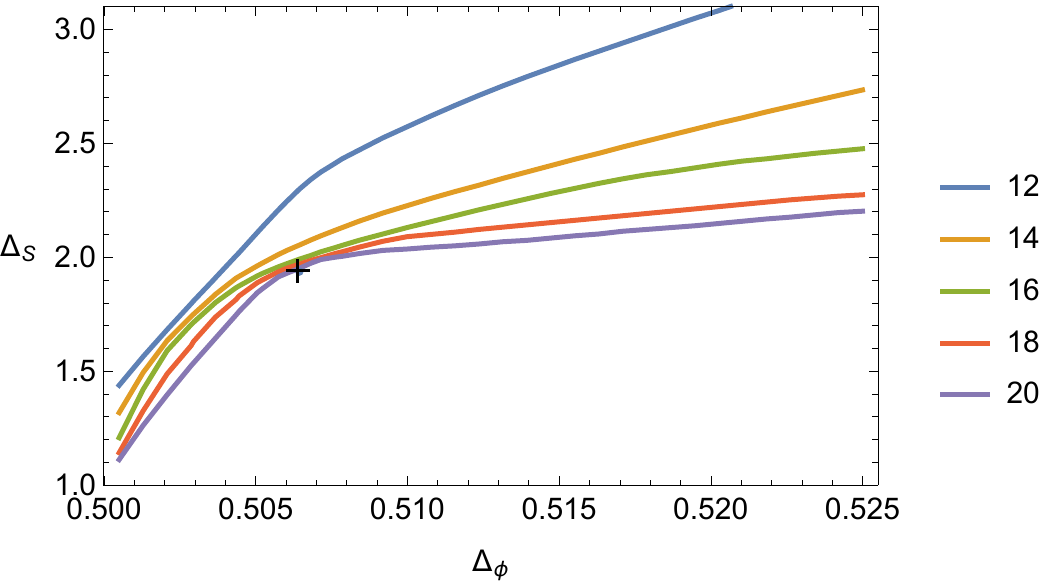}
\end{center}
\vspace*{-0.2cm}
\caption{\label{fig:7} 
\small
Bounds on  $\Delta_S$ as a function of $\Delta_\phi$ for $N=100$ points and different values of $\Delta_*$ for 3D CFTs with $O(20)$ symmetry, with $\phi$ in the fundamental representation of $O(20)$. The regions above the lines are excluded. 
The black cross marks the values of $\Delta_\phi$ and $\Delta_S$ for the $O(20)$ vector model as given in ref.~\cite{Kos:2013tga}. The curves and the labels in the legend have the same order from top to bottom.}
\begin{center}
\hspace*{0cm} 
\includegraphics[width=110mm]{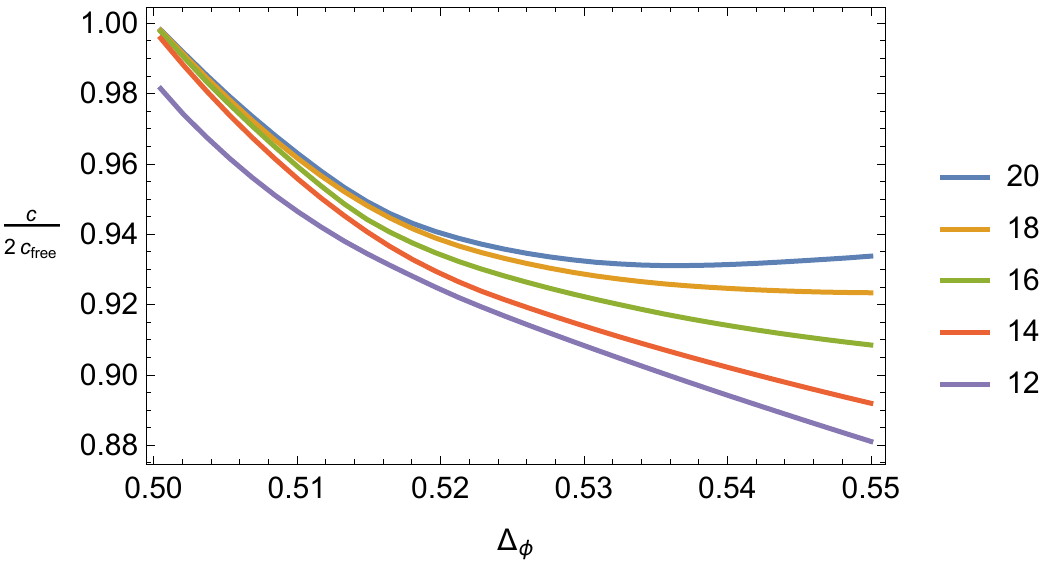}
\end{center}
\vspace*{-0.2cm}
\caption{\label{fig:8} 
\small
Bounds on  the central charge $c$  as a function of $\Delta_\phi$ for $N=100$ points and different values of $\Delta_*$  for 3D CFTs with $O(2)$ symmetry, with $\phi$ in the fundamental representation of $O(2)$.
Gaps $\Delta_S>1$ and $\Delta_T>1$ 
are assumed. The regions below the lines are excluded. The curves and the labels in the legend have the same order from top to bottom.}
\end{figure}

Let us now turn to 3D CFTs with $O(n)$ global symmetry. We consider a primary operator $\phi$ in the fundamental representation and denote the lowest-dimensional scalar singlet operator in the $\phi \times \phi$ OPE by $S$. It was found in refs.~\cite{Kos:2013tga,Kos:2015mba} that these CFTs have kinks in the bound on $\Delta_S$ as a function of $\Delta_\phi$ similar to that found for the Ising model. Moreover, the kinks coincide, for all values of $n$ that have been studied, with the values of $\Delta_\phi$ and $\Delta_S$ associated with the 
3D $O(n)$ models. On the other hand, a minimum in $c$ no longer occurs
for generic $O(n)$ models and the lower bound on $c$ instead monotonically decreases for $n>3$ (see ref.\cite{Kos:2013tga} for details).

In figs.~\ref{fig:5} and \ref{fig:6}, we show respectively the bound on $\Delta_S$ and $c$ (the latter normalized to the central charge $n c_{\rm free}$ of $n$ free scalars) as a function of $\Delta_\phi$ for different $O(n)$ symmetries, at fixed $N=80$ and $\Delta_*=16$.  For the central charge, gaps $\Delta_S>1$ and $\Delta_T>1$ in the spectrum of respectively singlet operators $S$ and rank-two symmetric-traceless operators $T$ are assumed as in ref.\cite{Kos:2013tga}. This assumption is satisfied for the $O(n)$ models and leads to more stringent bounds.
The dashed line corresponds to the leading large-$n$ prediction. 
All the qualitative behaviours found in ref.~\cite{Kos:2013tga} are reproduced, though with milder bounds, as expected.\footnote{Note however that no assumption on the spectrum
was made for the bounds on $\Delta_S$ presented in fig.~\ref{fig:5}, in contrast to fig.~2 of ref.~\cite{Kos:2013tga} where $\Delta_T>1$ was assumed.}
In particular,  the kinks in the $(\Delta_\phi$-$\Delta_S$) plane are not well visible at $\Delta_*=16$.
In figs.~\ref{fig:7} and \ref{fig:8}, we show the same bounds on $\Delta_S$ and $c$ as a function of $\Delta_\phi$ at fixed $N$ and $n$, for different values of $\Delta_*$.  We see the same qualitative behaviours regarding the ``UV sensitivities" found for 3D CFTs with no global symmetry (the Ising model). 
In particular, in fig.~\ref{fig:7} we see how the kink in the bound becomes well visible at $\Delta_*=18$  and does not significantly improve for $\Delta_*=20$. Its location is in very good agreement with that found in ref.~\cite{Kos:2013tga}.
On the other hand, the central-charge bound in fig.~\ref{fig:8} is still monotonically decreasing for $\Delta_*=18$ and a minimum appears only for $\Delta_*=20$. There are no signs of convergence comparing the bounds at $\Delta_*=18$ and $20$, indicating the need to go to larger $\Delta_*$ to approach the optimal bound.

\subsection{4D CFTs}
\label{subsec:4D}

\begin{figure}[!t]
\begin{center}
\hspace*{0cm} 
\includegraphics[width=110mm]{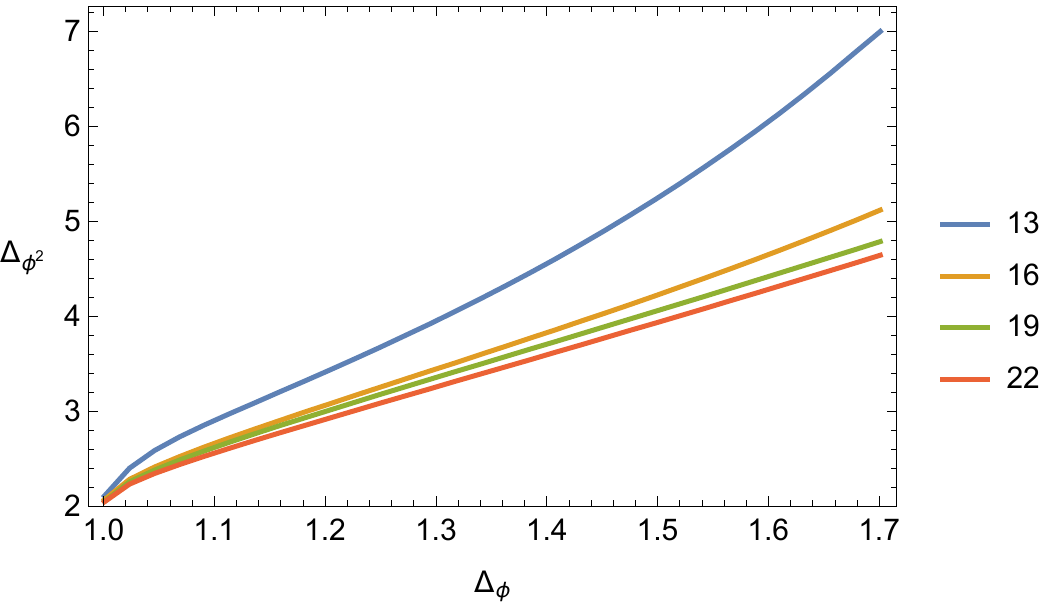}
\end{center}
\vspace*{-0.2cm}
\caption{\label{fig:4d_1} 
\small
Bounds on  $\Delta_{\phi^2}$ as a function of $\Delta_\phi$ for $N=100$ points and different values of $\Delta_*$ for 4D CFTs with no global symmetry. The regions above the curves are excluded. The curves and the labels in the legend have the same order from top to bottom.
}
\begin{center}
\hspace*{0cm} 
\includegraphics[width=110mm]{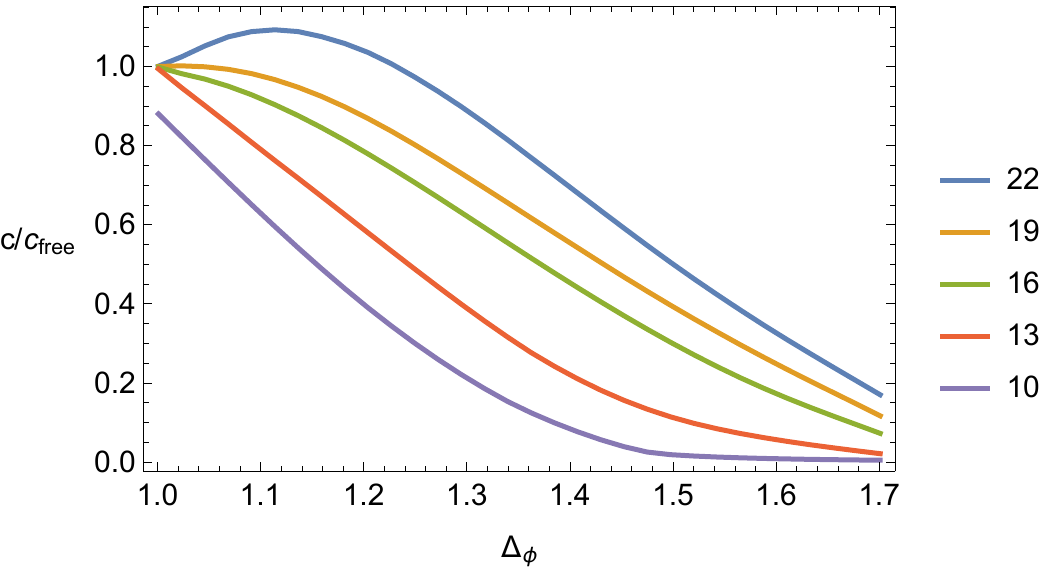}
\end{center}
\vspace*{-0.2cm}
\caption{\label{fig:4d_2} 
\small
Bounds on the central charge $c$ as a function of $\Delta_\phi$ for $N=100$ points and different values of $\Delta_*$ for 4D CFTs with no global symmetry.  The regions below the curves are excluded. The curves and the labels in the legend have the same order from top to bottom.
}
\end{figure}

All the above considerations can be repeated for 4D CFTs. There are no known non-super-symmetric CFTs at benchmarks points 
but it is still interesting to study
general bounds on operator dimensions and OPE coefficients. See e.g.~refs.~\cite{Rattazzi:2008pe,Caracciolo:2009bx,Poland:2011ey,Poland:2010wg, Rattazzi:2010gj, Rattazzi:2010yc,Caracciolo:2014cxa,Iha:2016ppj,Nakayama:2016knq}, 
where bounds of this kind (and others) have been determined
with the derivative method using both linear and semi-definite programming.

In figs.~\ref{fig:4d_1} and \ref{fig:4d_2}, we show bounds respectively on the dimension $\Delta_{\phi^2}$ of the lowest-dimensional scalar operator in the $\phi \times \phi$ OPE and on the central charge $c$ as a function of $\Delta_\phi$ for different values of $\Delta_*$, at fixed $N$. 
The conclusions are the same as for the 3D CFTs: the bounds on the operator dimension converge faster than those on the central charge.
The point of convergence of the bounds in $N$ at fixed $\Delta_*$ is again $N_*\sim \mathcal{O}(100)$ and thus also very similar to that in 3D CFTs.

\begin{figure}[!t]
\begin{center}
	\hspace*{0cm} 
	\includegraphics[width=110mm]{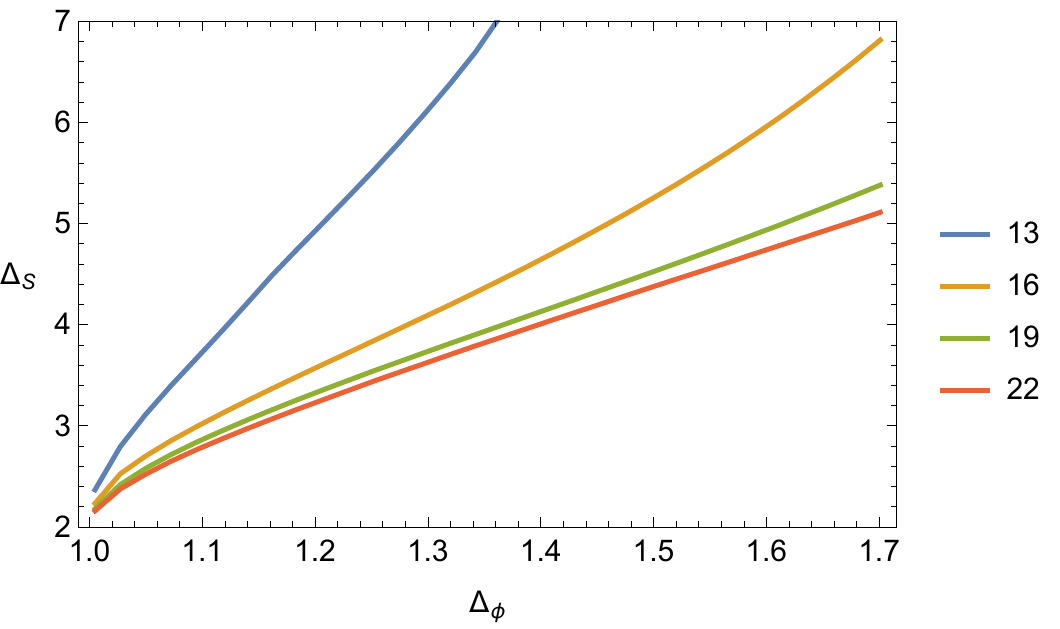}
\end{center}
\vspace*{-0.2cm}
\caption{\label{fig:4d_4} 
\small
Bounds on  $\Delta_S$ as a function of $\Delta_\phi$ for $N=100$ points and different values of $\Delta_*$ for 4D CFTs with $O(4)$ symmetry, with $\phi$ in the fundamental representation of $O(4)$. The regions above the curves are excluded. The curves and the labels in the legend have the same order from top to bottom.
}
\end{figure}

The analysis of 4D CFTs with $O(n)$ global symmetry also closely resembles its 3D counterpart. 
We again take the external field $\phi$ to transform in the fundamental representation of $O(n)$ and denote by $S$ the lowest-dimensional singlet scalar operator that appears in the $\phi\times\phi$ OPE.
For illustration, we report in fig.~\ref{fig:4d_4} the bound on $\Delta_S$ as a function of $\Delta_\phi$ for CFTs with $O(4)$ symmetry, at fixed $N$ and for different values of  $\Delta_*$. 
By comparing figs.\ref{fig:4d_1} and \ref{fig:4d_4} we notice that the convergence in $\Delta_*$ of the operator-dimension bound in 4D CFTs with $O(4)$ symmetry is slower than its analogue with no global symmetry.

\subsection{A Closer Look at the Spectrum of 3D $O(n)$ Models}
\label{subsec:spectrumON}

In the last subsections, we have shown how previously determined bounds are reproduced using the multipoint method.
Here we present some new results for the spectrum of $O(n)$ models. 
To this end we assume, as previous analyses indicate, that the 3D $O(n)$ models sit precisely at the kink on the boundary of the excluded region in the ($\Delta_\phi$-$\Delta_S$) plane ($\Delta_S$-maximization).
The vector $\vec{\rho}$ that we obtain from solving the linear program \eqref{LinearProgram} then gives us the spectrum and OPE coefficients of the operators that are exchanged in the $\langle \phi\phi\phi\phi\rangle$ correlator of the $O(n)$ models. 
Here we report the scaling dimensions of the first two
operators in respectively the singlet and rank-two representation of $O(n)$, $S$, $S^\prime$ and $T$, $T^\prime$, for $n=2,3,4$. Scalar operators with larger scaling dimensions are physically uninteresting, whereas 
$S^\prime$ and $T^\prime$ are important in determining the stability of the fixed points of the $O(n)$ models (being marginal operators in the underlying UV 4D Landau-Ginzburg theory) \cite{Calabrese:2002bm}.\footnote{See ref.\cite{Nakayama:2014lva} for a bootstrap approach to the 
study of the stability of fixed points in 3D $O(n)\times O(m)$ models.}
Actually, one additional operator should be considered, denoted as ${\cal P}_{4,4}$ in ref.\cite{Calabrese:2002bm}, but it transforms in the rank-four representation of $O(n)$ and hence cannot appear in the OPE of two scalar operators $\phi$ in the fundamental representation. Its dimension might be bounded (or computed) by considering a correlator involving, e.g., four $T$'s.
As far as we know, the scaling dimensions of $S^\prime$ and $T^\prime$ have not been
previously determined using the conformal bootstrap. The best determinations of these parameters have been made using a five-loop computation in the $\epsilon$-expansion in refs.\cite{Guida:1998bx} and \cite{Calabrese:2002bm}.\footnote{More precisely, $\Delta_{S^\prime}$ has been determined also by other means, such as fixed-dimension expansion and Monte Carlo simulations.
On the other hand, since $\Delta_{T^\prime}$ has been determined only using the $\epsilon$-expansion, we have decided to omit the other results for $\Delta_{S^\prime}$. The interested reader can find them, e.g., in table I of ref.\cite{Calabrese:2002bm}, where the coefficients $y_{4,0}$ and $y_{4,2}$ give $\Delta_{S^\prime} = 3-y_{4,0}$ and $\Delta_{T^\prime} = 3-y_{4,2}$. For completeness, we also report the relations
defining $\Delta_S$ and $\Delta_T$ in the notation of ref.\cite{Calabrese:2002bm}: $\Delta_S= 3-1/\nu$, $\Delta_T=3-y_{2,2}$.}

In table \ref{tableDelta}, we report the values of $\Delta_\phi$, $\Delta_S$, $\Delta_{S^\prime}$, $\Delta_T$, $\Delta_{T^\prime}$ determined in the literature,
for $n=2,3,4$. They should be compared with the values in table \ref{tableDeltaMultpoint} which have been determined in this paper as follows:
We take the values of $\Delta_\phi$ for $O(n)$ models with $n=2,3,4$ calculated in refs.~\cite{Campostrini:2006ms,Campostrini:2002ky,Hasenbusch:2000ph} as input and determine the scaling dimensions $\Delta_S$, $\Delta_{S'}$, $\Delta_T$ and $\Delta_{T'}$ using $\Delta_S$-maximization. We repeat this procedure for the lower, central and upper value of $\Delta_\phi$ given in these references and for different values of the cutoff $\Delta_*\in [18,23]$ and the number of points $N \in [60,120]$.\footnote{Our numerical precision does not allow us to take higher values of $\Delta_*$ and $N$ without having issues with numerical stability.} At fixed $N$ and $\Delta_*$,  we then take the average over the scaling dimensions obtained with the different input values of $\Delta_\phi$.  
\begin{table}[t]
\centering
\begin{tabular}{@{} c @{} c @{} c @{} c @{} c @{} c @{}}
\hline\hline
 $n$  &  \quad \quad $\Delta_\phi$ \quad  \quad  \quad&  \quad $\Delta_S$   \quad\quad \quad & \quad $\Delta_{S^\prime}$  \quad \quad  \quad &  \quad\quad $\Delta_T$ \quad  \quad \quad &   \quad $\Delta_{T^\prime}$ \quad  \quad \quad \\
\hline\hline
 2  \quad &  \quad 0.51905(10) \cite{Campostrini:2006ms}   \quad &\;  $1.5118^{+0.0012}_{-0.0022}$  \cite{Kos:2013tga} \quad & \; $3.802(18)$  \cite{Guida:1998bx}  &  $1.23613^{+0.00058}_{-0.00158}$ \cite{Kos:2013tga}& $3.624(10)$  \cite{Calabrese:2002bm}   \\ \hline
3   \quad &  \quad 0.51875(25) \cite{Campostrini:2002ky}   \quad &\; $1.5942^{+0.0037}_{-0.0047}$  \cite{Kos:2013tga} \quad & \; $3.794(18)$  \cite{Guida:1998bx}  \quad & \; $1.2089^{+0.0013}_{-0.0023}$ \cite{Kos:2013tga}& \; $3.550(14)$  \cite{Calabrese:2002bm}  \quad  \\
  \hline
4 \quad &  \quad 0.51825(40) \cite{Hasenbusch:2000ph}   \quad &  \; $1.6674^{+0.0077}_{-0.0087}$  \cite{Kos:2013tga} \quad & \; $3.795(30)$  \cite{Guida:1998bx}  \quad &  \;  $1.1864^{+0.0024}_{-0.0034}$ \cite{Kos:2013tga}& \; $3.493(14)$  \cite{Calabrese:2002bm}  \quad  \\
\hline
\end{tabular}
\caption{\small Scaling dimensions of the first two scalar operators in the singlet ($S$, $S^\prime$) and rank-two symmetric ($T$, $T^\prime$) representations of $O(n)$ for $n=2,3,4$ determined in the
literature.} 
\label{tableDelta}
\end{table}
\begin{table}[t]
\centering
\begin{tabular}{@{} c @{} c @{} c @{} c @{} c @{} c @{}}
\hline\hline
 \quad \quad  $n$  \quad \quad &  \quad \quad \quad $\Delta_\phi$ \quad  \quad  \quad&  \quad \quad\quad $\Delta_S$   \quad\quad \quad & \quad  \quad \quad $\Delta_{S^\prime}$  \quad \quad  \quad &  \quad\quad \quad $\Delta_T$ \quad  \quad \quad &  \quad \quad$\Delta_{T^\prime}$ \quad  \quad \quad \\
\hline\hline
 \quad \quad 2  \quad \quad &  \quad 0.51905(10) \cite{Campostrini:2006ms}   \quad&  \quad 1.5124(10)  \quad & 3.811(10) \quad &  \quad 1.2365(16)  & 3.659(7)  \quad   \\ 
 \hline
  \quad \quad 3  \quad \quad &  \quad 0.51875(25) \cite{Campostrini:2002ky}  \quad&  \quad 1.5947(35)  \quad & 3.791(22) \quad &  \quad 1.2092(22) & 3.571(12)   \quad   \\ 
  \hline
  \quad \quad 4  \quad \quad &  \quad 0.51825(40) \cite{Hasenbusch:2000ph}   \quad&  \quad 1.668(6)  \quad & 3.817(30)  \quad &  \quad 1.1868(24) & 3.502(16)  \quad   \\ 
\hline
\end{tabular}
\caption{\small Scaling dimensions of the first two scalar operators in the singlet ($S$, $S^\prime$) and rank-two symmetric ($T$, $T^\prime$) representations of $O(n)$ for $n=2,3,4$ determined in this paper using $\Delta_S$-maximization,  the values of $\Delta_\phi$ previously determined in the literature (first column) and the fit procedure explained in the main text.  The quoted error corresponds to 1$\sigma$ (68\% confidence level).}
\label{tableDeltaMultpoint}
\end{table}
\begin{table}[t]
\centering
\begin{tabular}{@{} c @{} c @{} c @{} c @{} c @{} c @{}}
\hline\hline
 \quad \quad  $n$  \quad \quad &  \quad \quad \quad $\Delta_\phi$ \quad  \quad  \quad&  \quad \quad\quad $\Delta_S$   \quad\quad \quad & \quad  \quad \quad $\Delta_{S^\prime}$  \quad \quad  \quad &  \quad\quad \quad $\Delta_T$ \quad  \quad \quad &  \quad  \quad$\Delta_{T^\prime}$ \quad  \quad \quad \\
\hline\hline
 \quad \quad 2  \quad \quad &  \quad 0.51905(10) \cite{Campostrini:2006ms}   \quad&  \quad $\leq$ 1.5145 \quad & $\leq$ 3.852 \quad &  \quad $\leq$ 1.2408 & $\leq$ 3.678  \quad   \\ 
 \hline
  \quad \quad 3  \quad \quad &  \quad 0.51875(25) \cite{Campostrini:2002ky}  \quad&  \quad  $\leq$ 1.6004 \quad & $\leq$ 3.856 \quad &  \quad $\leq$ 1.2116 & $\leq$ 3.588  \quad   \\ 
  \hline
  \quad \quad 4  \quad \quad &  \quad 0.51825(40) \cite{Hasenbusch:2000ph}   \quad&  \quad $\leq$ 1.677  \quad & $\leq$ 3.908 \quad &  \quad $\leq$ 1.191 & $\leq$ 3.528  \quad   \\ 
\hline
\end{tabular}
\caption{\small Upper bounds on the scaling dimensions of the first two scalar operators in the singlet ($S$, $S^\prime$) and rank-two symmetric ($T$, $T^\prime$) representations of $O(n)$ for $n=2,3,4$ determined in this paper
using $\Delta_S$-maximization and the values of $\Delta_\phi$ previously determined in the literature (first column).}
\label{tableMultpointUB}
\end{table}
Sometimes the same operator appears twice in the spectrum, at two different but close values of the scaling dimension. In this case we take the average of these values, weighted by the size of the corresponding OPE coefficient. Let us denote the resulting scaling dimensions by $\Delta_\mathcal{O}(N,\Delta_*)$ for $\mathcal{O}=S,S',T,T'$. Each of these values is associated with an error, resulting from the averaging.
The stepsize $\Delta_{{\rm step}}$ of our discretization has been set to $10^{-4}$ in the region where the operators were expected to be found 
(the resulting uncertainty in the scaling dimensions is typically negligible compared to the other errors).

\begin{figure}[!t]
\begin{center}
	\hspace*{0cm} 
	\includegraphics[width=100mm]{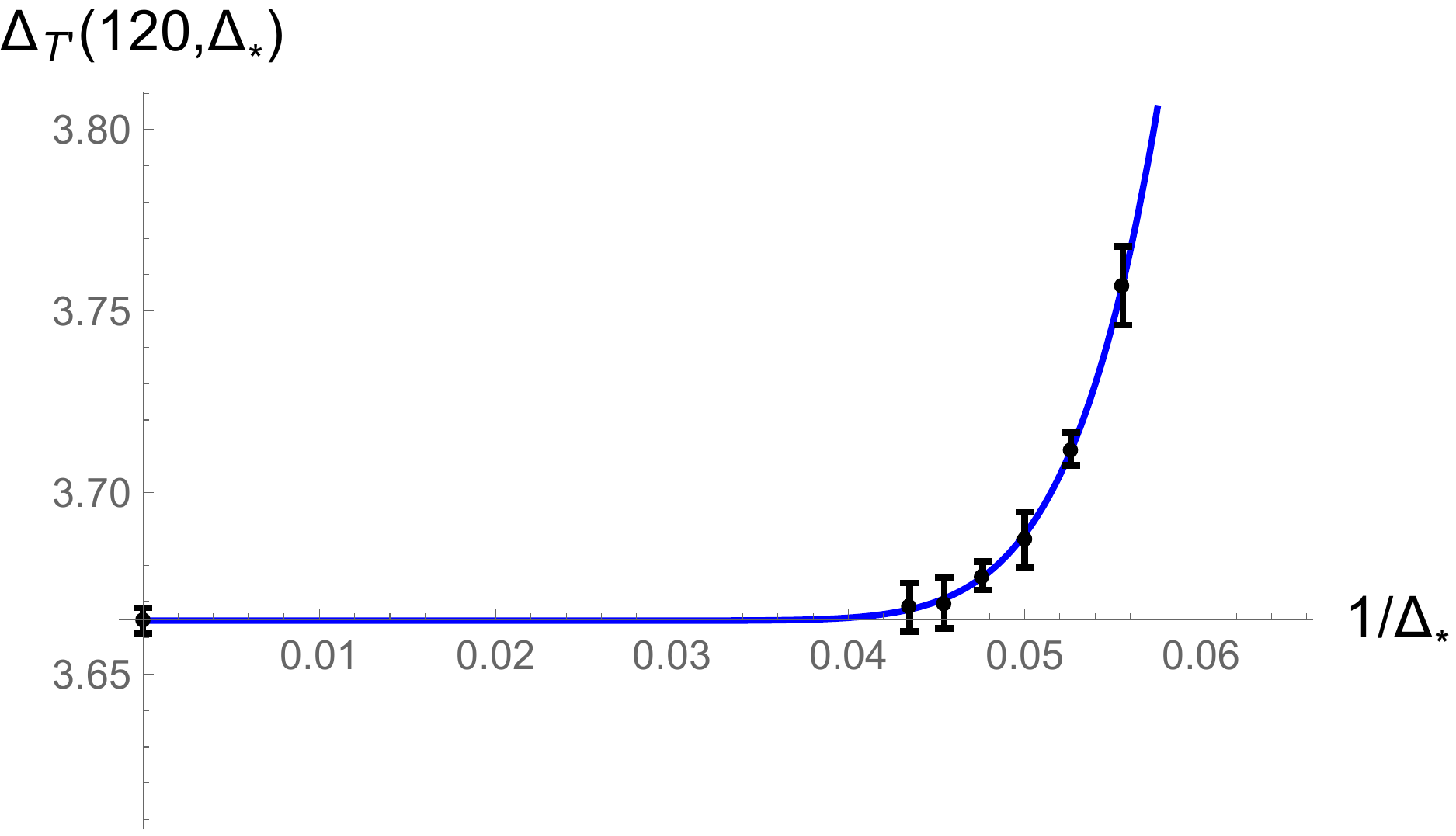}\\
\end{center}
\vspace*{-0.2cm}
\caption{\label{fig:a} 
\small
Extrapolation fit to determine the scaling dimension of the operator $T'$ in the $O(2)$ model with $N=120$ points at $\Delta_*=\infty$ from the results for that scaling dimension for different values of $\Delta_*$. The vertical error bar associated with the extrapolated point on the left corresponds to 1$\sigma$ (68\% confidence level).}
\begin{center}
	\hspace*{0cm} 
	\includegraphics[width=100mm]{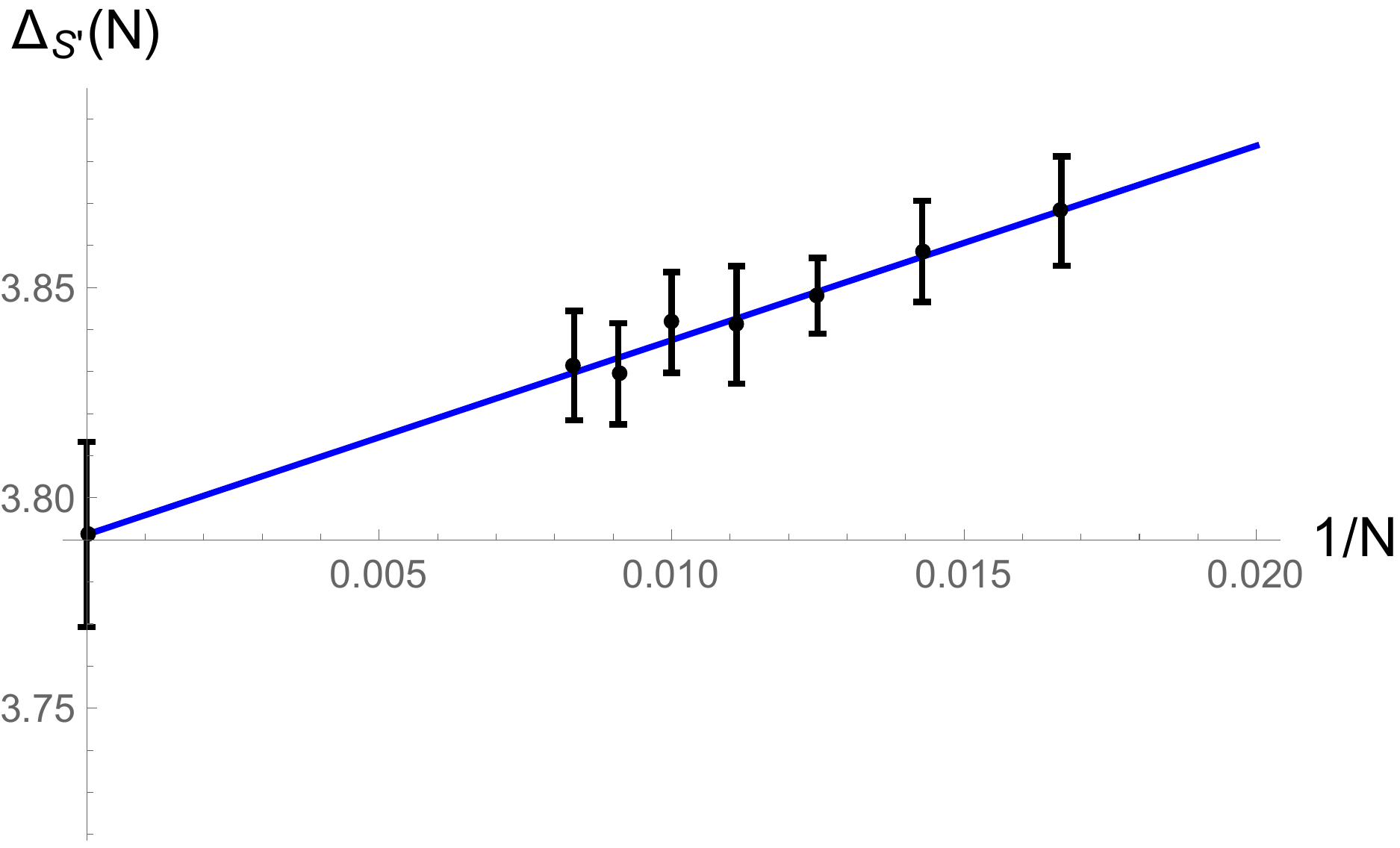}\\
\end{center}
\vspace*{-0.2cm}
\caption{\label{fig:b} 
\small
Extrapolation fit to determine the scaling dimension of the operator $S'$ in the $O(3)$ model at $N=\infty$ from the results for that scaling dimension for different values of $1/N$. Each point corresponds to the value of $\Delta_{S'}(N)$ extracted from a fit in $1/\Delta_*$. The vertical error bar associated with each point corresponds to 1$\sigma$ (68\% confidence level).}
\end{figure}

At fixed $N$, the results for different values of $\Delta_*$ are fitted by a function of the form $a_{\cal O}(N)+b_{\cal O}(N) \exp(- c_{\cal O}(N) \Delta_*)$, where $a_{\cal O}(N)$, $b_{\cal O}(N)$ and $c_{\cal O}(N)$ are the fit parameters. Such a dependence is roughly expected given the exponential convergence of the OPE.
Somewhat surprisingly, this simplified function fits the results extremely well, see fig.~\ref{fig:a} for an example of the extrapolation fit in $1/\Delta_*$.
Using this fit, we have extrapolated the scaling dimensions for the different operators and values of $N$ to $\Delta_*=\infty$. We denote the resulting scaling dimensions as $\Delta_{\cal O}(N)\equiv \Delta_{\cal O}(N,\infty) = a_{\cal O}(N)$.

We have then extrapolated to $N=\infty$ using a linear fit in $1/N$ which seems to well describe the behaviour of $\Delta_{\cal O}(N)$  as a function of $1/N$. An example of this extrapolation fit is shown in fig.\ref{fig:b}. We denote the resulting scaling dimensions as $\Delta_{{\cal O}} \equiv\Delta_{\cal O}(\infty)$.\footnote{A similar linear dependence in $1/N$ has already been noticed with great accuracy in ref.\cite{Beem:2015aoa} for the central-charge bound in 6D ${\cal N}=(2,0)$ SCFTs (see their fig.~1).} We do not have an analytic understanding of why the results should scale as $1/N$ for parametrically large $\Delta_*$. We simply take it as a working hypothesis. We expect that possible deviations from the linear behaviour
should be contained within the errors of our determination (cf.~fig.\ref{fig:b}).
Note that having $N$ as large as possible is clearly important for high precision. However, at fixed $\Delta_*$ the bounds saturate for sufficiently high $N$ and there is no gain in taking $N$ larger.

We have noticed that, at least for $n=2,3,4$, $\Delta_{\cal O}(N,\Delta_*)$ decreases as $N$ and/or $\Delta_*$ increase (this is obvious for $S$, but not for the other operators).
If we assume that this is true for any $N$ and $\Delta_*$, we may then set rigorous upper bounds without using any fit extrapolation. These bounds are reported in table \ref{tableMultpointUB}.
Comparing them with the results in table \ref{tableDeltaMultpoint} gives an idea of the impact of the fit extrapolation on the final results. 
As can be seen, all the scaling dimensions that we have determined are compatible with previous results in the literature. The only exception is $\Delta_{T'}$ for the $O(2)$ model for which our result has an approximate $3\sigma$ tension with that of ref.~\cite{Calabrese:2002bm}. Our accuracy in the determinations of $\Delta_{S}$ and $\Delta_{T}$ is comparable with that achieved in ref.~\cite{Kos:2013tga}, though it should be emphasized that the results there do not rely on extrapolations. Furthermore, our accuracy in the determinations of $\Delta_{S^\prime}$ and $\Delta_{T^\prime}$
is comparable with that achieved using the five-loop $\epsilon$-expansion. This is an indication that a slightly more refined bootstrap analysis will be able to improve the determinations of these scaling dimensions.

As we mentioned at the beginning of this subsection, $\Delta_S$-maximization also allows us to determine the OPE coefficients $\lambda_{\phi \phi {\cal O}}$.  We have not performed a detailed analysis with fit extrapolations as above to determine the asymptotic values of $\lambda_{\phi \phi {\cal O}}$ as $\Delta_*,N\rightarrow \infty$.
Instead we just report $\lambda_{\phi\phi S}$ as determined with the highest values $\Delta_*=22,23$ and $N=110,120$ used in this paper:
\bea
 O(2): \quad\, & \lambda_{\phi\phi S} & \approx 0.686  \,, \nn \\
O(3): \quad\, & \lambda_{\phi\phi S} & \approx 0.524 \,, \label{OPEcoeff} \\
O(4): \quad\, & \lambda_{\phi\phi S} & \approx 0.428  \,. \nn 
\eea
We have not determined the error associated with these results and have instead rounded them to the last shown digit. 
The results for $O(2)$ and $O(3)$ are in agreement with the recent determination in ref.\cite{Kos:2016ysd}, whereas the result for $O(4)$ is new as far as we know.

\section{Details of the Implementation}
\label{subsec:technicaldetails}

For the conformal blocks in $d=4$ dimensions, we use the closed-form expression from ref.~\cite{Dolan:2000ut}, normalized as in ref.~\cite{Rattazzi:2010yc}. For $d=3$ dimensions, on the other hand, we use the recursion relation for the conformal blocks found in 
ref.~\cite{Kos:2013tga}.\footnote{Alternatively, we can use the recursion relation also in $d=4$ dimensions by setting $d=4+\epsilon$ (to avoid double poles that appear at $d=4$). However, \texttt{Mathematica} evaluates the closed-form expression faster than (our implementation of) the recursion relation and we therefore choose the former.}
To this end, we iterate the recursion relation up to some cutoff $\Delta_{\rm rec}$. We choose this cutoff large enough such that the resulting error in the conformal blocks is smaller than the error from neglecting contributions of operators with dimensions larger than the truncation cutoff $\Delta_*$. In practice, we find that $\Delta_{\rm rec} = \Delta_* + \text{few}$ is sufficient to ensure this.

For the ansatz \eqref{lDeltavec} of discretized operator dimensions, we closely follow ref.~\cite{ElShowk:2012ht}. We generate the discrete spectra T1 to T4 (the latter only for sufficiently large $\Delta_*$) in their table 2, where we rescale the stepsizes $\delta$ by the factor $\Delta_{\rm step}/ (2 \cdot 10^{-5})$. We then remove duplicates from the combined spectrum and restrict to operator dimensions less than or equal to $\Delta_*$. We have performed extensive scans using different stepsizes $\Delta_{\rm step}$ and have found that the bounds converge for sufficiently small $\Delta_{\rm step}$. This is in particular satisfied for $\Delta_{\rm step}= 2\cdot 10^{-3}$ which we choose for all the plots in this paper. For the determination of the spectra in sec.~\ref{subsec:spectrumON} we add additional operators with stepsize  $\Delta_{\rm step}= 10^{-4}$ around the previously determined scaling dimensions for the operators  $S$, $S'$, $T$, $T'$ in the $O(n)$ models. Furthermore, for bounds on operator dimensions for which the plots extend to bounds $\Delta_{\phi^2}> 3$ (the largest dimension of T1 of ref.~\cite{ElShowk:2012ht}), we have included additional operators in the scalar sector so that the smallest stepsize $\Delta_{\rm step}$ is used up to the largest bound on $\Delta_{\phi^2}$ shown in that plot.
We have also performed scans using different parametrizations for the ansatz \eqref{lDeltavec} and have found that the bounds become indistinguishable from the bounds obtained with the ansatz discussed above for sufficiently small $\Delta_{\rm step}$. This gives us confidence that the discretization does not introduce any artifacts into our calculations.

We use \texttt{Mathematica} to evaluate the conformal blocks for the different operators that appear in the ansatz \eqref{lDeltavec} and for the set of points in the $z$-plane. The linear progam \eqref{LinearProgram} is then set up by a program written in Python and is subsequently solved with the optimizer \texttt{CPLEX} by \texttt{IBM} using the primal simplex algorithm. Since this optimizer is limited to double precision, it is important to reduce the spread in size of the numerical values in the problem. To this end, note that we can rescale each row of the inequality \eqref{LinearProgram} separately by a positive number. Denoting a given row by $\mathcal{R}$, we rescale its elements by
\be 
\mathcal{R}^{\rm resc}_i \, = \, \frac{\mathcal{R}_i}{\sqrt{\underset{i}{\min} |\mathcal{R}_i| \cdot \underset{i}{\max} |\mathcal{R}_i|}} \, .
\ee
Similarly, we can rescale each column of the matrix $\mathcal{M}$ separately by a positive number if we redefine the corresponding (squared) OPE coefficient in the vector $\vec{\rho}$. We again choose
\be 
\mathcal{M}^{\rm resc}_{ij} \, = \, \frac{\mathcal{M}_{ij}}{\sqrt{\underset{i}{\min} |\mathcal{M}_{ij}| \cdot \underset{i}{\max} |\mathcal{M}_{ij}|}}
\ee
and correspondingly for $\vec{\rho}$. This procedure is iterated three times in our Python code, using precision arithmetric with 120 digits to ensure that no significant rounding errors are introduced in the process (the conformal blocks have been calculated with the same precision). Since we perform our own rescaling, we switch off this option in \texttt{CPLEX}. 

We find that the above rescaling typically reduces the orders of magnitude in the ratio between the largest and smallest numerical value in eq.~\eqref{LinearProgram} by about half. Nevertheless, precision is a limiting factor and does not allow us to go to cutoffs $\Delta_*$ much larger than 20. The fact that double precision is sufficent for smaller cutoffs, on the other hand, makes our calculations (combined with the excellent speed of \texttt{CPLEX}) very fast.

\section{Conclusions}
\label{conclusions}

We have implemented the method proposed in ref.\cite{Hogervorst:2013sma} to numerically study the bootstrap equations away from the symmetric point $z=\bar z=1/2$.
Using this method, we have qualitatively reproduced various results that have been determined in the bootstrap literature using the more common method of taking derivatives at the symmetric point.
The main aim of our work was to show that bootstrapping with multipoints works and is a valid alternative to the standard derivative method.
In particular, it can be useful at a preliminary stage when one wants to qualitative bound or approximately compute some quantities using the bootstrap. By choosing
a sufficiently low cutoff $\Delta_*$, one can get qualitatively good results within seconds of CPU time with a standard laptop!
Since the optimizer \texttt{CPLEX} that we use is limited to double precision, we can not achieve the high precision of refined bootstrap codes such as Juliboots \cite{Paulos:2014vya} or SDPB \cite{Simmons-Duffin:2015qma}. Nevertheless we have shown how, using $\Delta$-maximization, relatively precise results can be obtained for the scaling dimensions of operators (though we relied on an extrapolation procedure). In particular, for $O(n)$ models with $n=2,3,4$ we have determined the scaling dimensions of the second-lowest-dimensional operators $S'$ and $T'$ in the singlet and symmetric-traceless representation, respectively. To our knowledge, these have not been determined before using bootstrap techniques. We believe that it should not be difficult to go to arbitrary precision and get rid of the discretization (and the extrapolation procedure) by, for instance, adapting the algorithm developed in refs.\cite{El-Showk:2014dwa,Paulos:2014vya} to multipoints. We do not exclude that bootstrapping with multipoints might then turn out to be comparable to (or better than) the derivative method for high-precision
computations. From a conceptual point of view, the multipoint method is more rigorous, since the crossing equations are not truncated but bounded by an error.\footnote{Strictly speaking, this is true only when we are guaranteed to be in the regime where the Hardy-Littlewood tauberian theorem applies. But all the evidence so far indicates that this is always the case for $\Delta_*\gtrsim O(10)$.}

We have also discussed how the multipoint method is useful in understanding to which extent a given numerical result depends sensitively on the high-dimensional operators.
In particular, we have noticed that bounds on operator dimensions are less sensitive in this respect than bounds on the central charge.  

Ideally, one might want to push the multipoint method to the extreme ``IR limit", by choosing a cutoff $\Delta_*$ so low that an analytic approach may become possible.
This is certainly a very interesting direction that should be explored. Among other things, it requires to improve on the estimate of the OPE convergence given in ref.\cite{Pappadopulo:2012jk} that applies in the opposite regime, for parametrically large $\Delta_*$. Perhaps the results of ref.\cite{Kim:2015oca} might be useful in this respect.\footnote{It should be mentioned that another method based on  determinants has been proposed by Gliozzi  \cite{Gliozzi:2013ysa} to study severely truncated bootstrap equations. The method sometimes works surprisingly well, but unfortunately its current implementation is not systematic and rigorous enough.} 

An important line of development in the numerical bootstrap is the analysis of mixed correlators which so far are numerically accessible only using semi-definite programming \cite{Kos:2014bka}. It would be very interesting to implement mixed correlators in the multipoint bootstrap, either by adapting the semi-definite programming techniques or by extending the linear programming techniques.

\section*{Acknowledgments}

M.S. thanks Pasquale Calabrese, Sheer El-Showk, Miguel F. Paulos, Alessandro Vichi and especially Slava Rychkov for useful discussions.
We also thank Slava Rychkov for comments on the manuscript.
We warmly acknowledge Antonio Lanza, Piero Calucci and all members of SISSA - ITCS (Information Technology and Computing Services) for their continuous  availability and support 
in using the Ulysses HPC facility maintained by SISSA.
The work of M.S. was supported by the ERC Advanced Grant no. 267985 DaMESyFla. A.C.E. and M.S. gratefully acknowledge
support from the Simons Center for Geometry and Physics at Stony Brook University, 
where some of the research for this paper was performed.

\end{document}